\documentclass[12pt,letterpaper]{article}

\usepackage{xr}
\makeatletter
\newcommand*{\addFileDependency}[1]{% argument=file name and extension
  \typeout{(#1)}
  \@addtofilelist{#1}
  \IfFileExists{#1}{}{\typeout{No file #1.}}
}
\makeatother
 
\newcommand*{\myexternaldocument}[1]{%
    \externaldocument{#1}%
    \addFileDependency{#1.tex}%
    \addFileDependency{#1.aux}%
}

\myexternaldocument{vecchia_laplace_supp}
\usepackage{mathtools}
\mathtoolsset{showonlyrefs} 

\usepackage{natbib}
\usepackage[ left=1in, top=1in, right=1in, bottom=1in]{geometry}
\usepackage{xcolor,graphicx,bm,colonequals,amsmath,amssymb,url}
\usepackage{array,tabularx,multirow}
\usepackage[shortlabels]{enumitem}
\usepackage[font={footnotesize}]{caption,subcaption}
\usepackage{siunitx}

\newcolumntype{d}[1]{D{.}{.}{#1}}
\usepackage{color}

\usepackage[normalem]{ulem} % %
\usepackage{bbm}
\newcommand{\indicat}{\mathbbm{1}}

\bibpunct[, ]{(}{)}{;}{a}{,}{,}

\usepackage{amsthm}
\newtheoremstyle{propstyle} % name
    {3mm}                    % Space above
    {1mm}                    % Space below
    {\itshape}                   % Body font
    {}                           % Indent amount
    {\scshape}                   % Theorem head font
    {.}                          % Punctuation after theorem head
    {.5em}                       % Space after theorem head
    {}  % Theorem head spec (can be left empty, meaning ‘normal’)
\theoremstyle{propstyle}

\theoremstyle{propstyle}

\theoremstyle{propstyle}

\newsavebox\ideabox

\newcommand{\ba}{\mathbf{a}}
\newcommand{\bh}{\mathbf{h}}

\newcommand{\bb}{\mathbf{b}}

\newcommand{\bs}{\mathbf{s}}

\newcommand{\bt}{\mathbf{t}}
\newcommand{\bx}{\mathbf{x}}
\newcommand{\bu}{\mathbf{u}}
\newcommand{\by}{\mathbf{y}}

\newcommand{\bz}{\mathbf{z}}

\newcommand{\bW}{\mathbf{W}}

\newcommand{\bD}{\mathbf{D}}

\newcommand{\bU}{\mathbf{U}}
\newcommand{\bV}{\mathbf{V}}
\newcommand{\bK}{\mathbf{K}}

\newcommand{\bQ}{\mathbf{Q}}

\newcommand{\bfzero}{\mathbf{0}}
\newcommand{\bfalpha}{\bm{\alpha}}

\newcommand{\bfmu}{\bm{\mu}}

\newcommand{\bftheta}{\bm{\theta}}

\DeclareMathOperator{\E}{E}

\DeclareMathOperator{\diag}{diag}
\DeclareMathOperator{\chol}{chol}
\DeclareMathOperator{\rchol}{rchol}
\DeclareMathOperator{\rev}{rev}
\DeclareMathOperator{\logit}{logit}

\newcommand{\GP}{GP}

\newcommand{\normal}{\mathcal{N}}
\newcommand{\order}{\mathcal{O}}

\newcommand{\domain}{\mathcal{D}}

\newcommand{\dens}{p}
\newcommand{\adens}{\widehat{p}}

\newcommand{\locs}{\mathcal{S}}

\DeclareMathOperator*{\argmax}{arg\,max}

\title{
Vecchia-Laplace approximations of generalized Gaussian processes for big non-Gaussian spatial data
}

\author{
Daniel Zilber\thanks{Department of Statistics, Texas A\&M University} 
\and 
Matthias Katzfuss\footnotemark[1] \thanks{Corresponding author: \texttt{katzfuss@gmail.com}}
}

\graphicspath{{plots/}}

\usepackage{algorithm, algpseudocode}
\begin{document}

\maketitle

\begin{abstract}
Generalized Gaussian processes (GGPs) are highly flexible models that combine latent GPs with potentially non-Gaussian likelihoods from the exponential family. GGPs can be used in a variety of settings, including GP classification, nonparametric count regression, modeling non-Gaussian spatial data, and analyzing point patterns. However, inference for GGPs can be analytically intractable, and large datasets pose computational challenges due to the inversion of the GP covariance matrix. We propose a Vecchia-Laplace approximation for GGPs, which combines a Laplace approximation to the non-Gaussian likelihood with a computationally efficient Vecchia approximation to the GP, resulting in simple, general, scalable, and accurate methodology. We provide numerical studies and comparisons on simulated and real spatial data. Our methods are implemented in a freely available R package.
\end{abstract}


%%%%%%%%%%%%%%%%%%%%%%%%%%%%%%%%%%%%%%%%%%%%%%%%%%%%%%%

\section{Introduction \label{sec:intro}}

% Brief definition of GGPs, including some potential application areas (with citations):
Dependent non-Gaussian data are ubiquitous in time series, geospatial applications, and more generally in nonparametric regression and classification. A popular model for such data is obtained by combining a latent Gaussian process (GP) with conditionally independent, potentially non-Gaussian likelihoods from the exponential family. This is traditionally referred to as a spatial generalized linear mixed model (SGLMM) in the spatial statistics literature \citep{Diggle1998}, but the same model has more recently also been referred to as a generalized GP \citep[GGP; e.g.,][]{chan2011generalized}; we will use the latter, more concise term throughout. GGPs are highly flexible, interpretable, and allow for natural, probabilistic uncertainty quantification. However, inference for GGPs can be analytically intractable, and large datasets pose additional computational challenges due to the inversion of the GP covariance matrix. 

% Summary of literature on approximation of the non-Gaussian likelihood:
Popular techniques to numerically perform the intractable marginalization necessary for inference are, in order of increasing speed: Markov chain Monte Carlo (MCMC), expectation propagation, variational methods, and Laplace approximations. See \citet{Shang2013} for a recent review of deterministic techniques and \citet{Filippone2014} for a comparison of MCMC and expectation propagation. \citet{Rue2009} argues that variational methods and expectation propagation suffer from underestimated and overestimated posterior variances, respectively. Here, we consider the Laplace approximation \citep[e.g.,][]{tierney1986accurate,williams1998bayesian}, a classic technique for integral evaluation based on second-order Taylor expansion. \citet{bonat2016practical} show numerically that the Laplace approximation can be a practical and accurate method for GGP inference.

% Brief summary of literature on approximation of GPs (with Gaussian noise) for large data --- combine approaches from statistics and ML literatures
It has long been recognized that the computational cost for GPs is cubic in the data size.
Much work has been done on GP approximations that address this problem in the context of Gaussian noise \citep[as recently reviewed in][]{Heaton2017}. Low-rank approaches \citep[e.g.,][]{Higdon1998, Quinonero-Candela2005, Banerjee2008, Cressie2008, Katzfuss2010} have limitations in the presence of fine-scale structure \citep{Stein2013a}, but they have proved popular due to their simplicity.  
Approximations relying on sparsity in covariance matrices \citep{furrer2006covariance,kaufman2008covariance} by definition can only capture local, short-range dependence and cannot guarantee low computation cost.  Approaches that take advantage of Toeplitz or Kronecker structure \citep[e.g.,][]{dietrich1997fast, flaxman2015fast, guinness2017circulant} can be extremely efficient but are not as generally applicable.
Recently, methods relying on sparsity in precision matrices \citep{rue2005gaussian,lindgren2011explicit,Nychka2012} have gained popularity due to their accuracy and flexibility. In particular, a class of highly promising GP approximations \citep{Vecchia1988,Stein2004,Datta2016,Guinness2016a,Katzfuss2017a,Katzfuss2018} rely on ordered conditional independence that can guarantee linear scaling in the data size while resolving dependence at all scales.

%More thorough review of the literature on approximation methods that address both non-Gaussianity and large $n$:
There are also a number of existing methods for large non-Gaussian datasets modeled using GGPs. A popular approach is to combine a low-rank GP with an approximation of the non-Gaussian likelihood, as the dimension reduction inherent in the low-rank approximation carries through even to the non-Gaussian case. \citet{Sengupta2013} estimate parameters using an expectation-maximization algorithm with low-rank and Laplace approximations. \citet{Sheth2015} use variational inference to obtain the posterior and select a set of conditioning points for their low-rank approximation. 
%Low-rank methods can be generalized with stochastic variational inference \citep{hensman2013gaussian} and linear projections onto lower dimensional manifolds \citep{banerjee2012efficient}. 
Some methods of dimension reduction, including random sketching \citep[e.g.,][]{yang2017randomized} and projection, offer theoretical guarantees and can be combined with MCMC methods for the analysis of non-Gaussian data \citep[e.g.,][]{hughes2013dimension, guan2018computationally}, but are still subject to the limitations of low-rank methods.
\citet{nickisch2018state} develop state-space models for one-dimensional non-Gaussian time series, which can perform inference in linear time and memory using a set of knots in time, a form of low-rank approximation. Alternate priors \citep[e.g., log gamma priors for count data,][]{bradley2018computationally} are an interesting but specialized approach to completely avoid the intractability issues with GGPs. 

% Important: straightforward combination of vecchia and Laplace (NNGP) is not fully scalable. No guarantee that sparsity can be maintained under all relevant operations
Similar to what we shall propose, some authors have combined a sparse-precision approach with a non-Gaussian approximation. A prominent example is \citet{lindgren2011explicit}, in which an integrated nested Laplace approximation (INLA) is combined with a sparse-precision approximation of the GP using its representation as the solution to a stochastic partial differential equation. \citet{Datta2016} proposed to apply the GP approximation of \citet{Vecchia1988} to a latent GP, but did not provide an explicit algorithm for large non-Gaussian data. While both \citet{lindgren2011explicit} and \citet{Datta2016} limit the number of nonzero entries per row or column in the precision matrix to a small constant, the computational complexity for decomposing this sparse $n \times n$ matrix is not linear in $n$, but rather $\order(n^{3/2})$ in two dimensions \citep[][Thm.~6]{Lipton1979}, and at least $\order(n^2)$ in higher dimensions. In the Gaussian setting, this scaling problem can be overcome by applying a Vecchia approximation to the observed data \citep{Vecchia1988} or to the joint distribution of the observed data and the latent GP \citep{Katzfuss2017a}.

% then description of our proposed VL, including a brief summary of its properties (linear complexity)
To handle both scaling and intractability, we propose a Vecchia-Laplace (VL) approximation for GGPs. The posterior mode necessary for the Laplace approximation is found using the Newton-Raphson algorithm, which can be viewed as iterative GP inference based on Gaussian pseudo-data. At each iteration of our VL algorithm, the joint Gaussian distribution of the pseudo-data and the latent GP realizations is approximated using the general Vecchia framework \citep{Katzfuss2017a,Katzfuss2018}. By modeling the joint distribution of pseudo-data and GP realizations at each iteration, our VL approach can ensure sparsity and guarantee linear scaling in $n$ for any dimension, overcoming the scaling issues of the sparse-matrix approaches mentioned above. 

% novelty and advantages
To our knowledge, we provide the first explicit algorithm extending and applying the highly promising class of general-Vecchia GP approximations to large non-Gaussian data.  We believe it to be a useful addition to the literature due to its speed, simplicity, guaranteed numerical performance, and wide applicability (e.g., binary, count, right-skewed, and point-pattern data).  In addition, as shown in \citet{Katzfuss2017a}, the general Vecchia approximation includes many popular GP approximations \citep[e.g.,][]{Vecchia1988,Snelson2007,Finley2009,Sang2011a,Datta2016,Katzfuss2015,Katzfuss2017b} as special cases, and so our VL methodology also directly provides extensions of these GP approximations to non-Gaussian data.

% paper outline:
The remainder of this document is organized as follows. In Section \ref{sec:review}, we review the Laplace approximation and general Vecchia.  In Section \ref{sec:vlapprox}, we introduce and examine our VL method, including parameter inference and predictions at unobserved locations. In Sections \ref{sec:simulation} and \ref{sec:application}, we study and compare the performance of VL on simulated and real data, respectively. Some details are left to the appendix. A separate Supplementary Material document contains Sections \ref{supp:vllikelihood}--\ref{supp:MODIS} with additional derivations, simulations, and discussion.
The methods and algorithms proposed here are implemented in the \texttt{R} package \texttt{GPvecchia} \citep{GPvecchia} with sensible default settings, so that a wide audience of practitioners can immediately use the code with little background knowledge. Our results and figures can be reproduced using the code and data at \url{https://github.com/katzfuss-group/GPvecchia-Laplace}.

%%%%%%%%%%%%%%%%%%%%%%%%%%%%%%%%%%%%%%%%%%%%%%%%%%%%%%%
\section{Review of existing results \label{sec:review}}

%%%%%
\subsection{Generalized Gaussian processes\label{sec:GGP}}

Let $y(\cdot) \sim \GP(\mu,K)$ be a latent Gaussian process with mean function $\mu$ and kernel or covariance function $K$ on a domain $\domain \subset \mathbb{R}^d$, $d \in \mathbb{N}^+$. Observations $\bz = (z_1,\ldots,z_n)'$ at locations  $\bs_i \in \domain$ are assumed to be conditionally independent, $z_i | \by \stackrel{ind.}{\sim} g_i(z_i|y_i)$, where $\by = (y_1,\ldots,y_n)'$ and $y_i = y(\bs_i)$. 
%When working with a vector of functions $\bF$, we denote the entrywise evaluation across a vector of values $\bv$ with: $$\bF_\bv = \left( F_1(v_1), F_2(v_2), ..., F_n(v_n)) \right)$$
We assume that the observation densities or likelihoods $g_i$ are from the exponential family. Parameters $\bftheta$ in $\mu$, $K$, or the $g_i$ will be assumed fixed and known for now; for example, $\bftheta$ may contain regression coefficients determining the mean function $\mu$, or variance, smoothness, and range parameters determining a Mat\'ern covariance $K$.

Our goal is to obtain an approximation of the posterior of $\by$, which takes the form    
\begin{equation}
\label{eq:exactposterior}
p(\by|\bz) = \frac{ \normal_n(\by|\bfmu,\bK) \, \prod_{i=1}^n g_i(z_i|y_i)}{p(\bz)},
\end{equation}
where $\bfmu = (\mu(\bs_1),\ldots,\mu(\bs_n))'$, and $\bK$ is an $n\times n$ covariance matrix with $(i,j)$ entry $(\bK)_{i,j} = K(\bs_i,\bs_j)$.
Once an approximation of the posterior \eqref{eq:exactposterior} has been obtained, it is conceptually straightforward to extend this result to other quantities of interest, such as the integrated likelihood for inference on parameters $\bftheta$ (see Section \ref{sec:likelihood}), and prediction of $y(\cdot)$ at unobserved locations (see Section \ref{sec:prediction}).

%%%%%
\subsection{Review of the Laplace approximation \label{sec:laplace}}

The normalizing constant $p(\bz)$ in \eqref{eq:exactposterior} is not available in closed form for non-Gaussian likelihoods. A popular approach to this issue is the Laplace approximation \citep[e.g.,][Section 3.4]{williams1998bayesian, Rasmussen2006}, which approximates $p(\bz) = \int \exp(\log p(\bz|\by)) p(\by) d\by$ via a second-order Taylor expansion of $\log p(\bz|\by)$ at the mode of the posterior density $p(\by|\bz)$. As this results in an exponentiated quadratic form in $\by$, it is equivalent to a Gaussian approximation of the likelihood.
The mode of $\log p(\by|\bz)$ does not depend on the normalizing constant, and so it can be obtained using standard optimization procedures such as the Newton-Raphson algorithm. The crucial observation for our later developments is that each Newton-Raphson update in the GGP setting is equivalent to computing the posterior mean of $\by$ given \emph{Gaussian} pseudo-data \citep[e.g.,][Sect.~3.4.1]{Rasmussen2006}. Upon convergence of the algorithm, we have a Laplace approximation for the normalizing constant and a Gaussian approximation for the likelihood, which gives us a Gaussian posterior. 

We now go into the details of this approximation. Based on the first and second derivative of $\log g_i$, we define 
\[
\textstyle u_i(y_i) = \frac{\partial}{\partial y_i} \log g_i(z_i|y_i) \qquad \text{and} \qquad d_i(y_i) = -\big(\frac{\partial^2}{\partial y_i^2} \log g_i(z_i|y_i)\big)^{-1}, \qquad i=1,\ldots,n.
\]
Stacking these quantities as
$\bu_\by = \big(u_1(y_1),\ldots,u_n(y_n)\big)'$ 
and
$\bD_\by = \diag\big(d_1(y_1),\ldots,d_n (y_n)\big)$,
it is easy to see that $\frac{\partial}{\partial \by} \log p(\by|\bz) =- \bK^{-1}(\by-\bfmu) + \bu_\by$ and $-\frac{\partial^2}{\partial \by \partial \by'} \log p(\by|\bz) = \bK^{-1} + \bD_\by^{-1} \equalscolon \bW_\by$.  When given the posterior mode $\bfalpha =\argmax_{\by \in \mathbb{R}^n} \log p(\by|\bz)$, the combined Gaussian/Laplace approximation of the posterior is
\begin{equation}
\label{eq:laplaceposterior}
\widehat p_L(\by | \bz) = \normal_n(\by|\bfalpha,\bW^{-1}_{\bfalpha}).
\end{equation}
The subscript $\bfalpha$ in $\bW^{-1}_{\bfalpha}$ implies evaluation of $\bW_\by$ at the mode $\bfalpha$, rather than at an arbitrary $\by$. To obtain the mode $\bfalpha$ with the Newton-Raphson algorithm, we start with an initial value $\by^{(0)}$, and update the current guess for $\ell=0,1,2,\ldots$ until convergence as $\by^{(\ell+1)} = \bh(\by^{(\ell)})$, where
\begin{equation}
    \label{eq:nr}
\textstyle \bh(\by) = \by - \big(\frac{\partial^2}{\partial \by \partial\by'} \log p(\by|\bz)\big)^{-1}\big(\frac{\partial}{\partial \by} \log p(\by|\bz)\big).
\end{equation}
This Newton-Raphson update is equivalent to computing the posterior mean of $\by$ given Gaussian pseudo-data $\bt_\by = \by + \bD_\by \bu_\by$  with noise covariance matrix $\bD_\by$.  Specifically, we can write the Newton-Raphson update in \eqref{eq:nr} as:
\begin{equation}
\label{eq:postmean}
\bh(\by) = \bfmu + \bW_\by^{-1}\bD_\by^{-1}(\bt_\by - \bfmu) = \E(\by|\bt_\by),
\end{equation}
which is the conditional mean of $\by$ given Gaussian pseudo-data $\bt_\by | \by \sim \normal_n(\by,\bD_\by )$.
%This fits naturally into the general Vecchia framework (Section \ref{sec:vecchia}), which models the data and latent variables jointly.  
The derivation of \eqref{eq:postmean} is straightforward and included in Appendix \ref{app:nr} for completeness. 
This means we can obtain the mode $\bfalpha$ by iterating between (a) computing pseudo-data $\bt_{\by^{(\ell)}}$ with $i$th entry $y_i^{(\ell)} + d_i(y_i^{(\ell)}) u_i(y_i^{(\ell)})$, and (b) obtaining the posterior mean $\by^{(\ell+1)}$ of $\by$ given $\bt_{\by^{(\ell)}}$ assuming independent Gaussian noise with variances $d_1(y_1^{(\ell)}),\ldots,d_n(y_n^{(\ell)})$.

\begin{table}
\centering
 \begin{tabular}{l | r | r | r } 
distribution & likelihood $g(z|y)$ & pseudo-data $t_y$ & pseudo-variance $d(y)$ \\ 
 \hline
Gaussian & $\normal(y,\tau^2)$ & $z$ & $\tau^2$ \\ 
 Bernoulli & $\mathcal{B}(\logit^{-1}(y))$ &$ y+\frac{(1+e^{y})^2}{e^y}(z - \frac{e^y}{1+e^y})  $& $(1+e^{-y})(1+e^y)$ \\ 
Poisson & $\mathcal{P}(e^y)$ & $y + e^{-y}(z -e^{y}) $   & $e^{-y}$ \\
Gamma & $\mathcal{G}(a, ae^{-y})$ & $y + (1 - z^{-1}e^y)$   & $aze^{-y}$ \\
\end{tabular}
\caption{Examples of popular likelihoods, together with the Gaussian pseudo-data and pseudo-variances implied by the Laplace approximation. The non-canonical logarithmic link function is used for the Gamma likelihood to ensure that the second parameter, $ae^{-y}$, is positive.}
\label{tab:likelihoods}
\end{table}

\begin{figure}
\centering
\includegraphics[width=.95\textwidth]{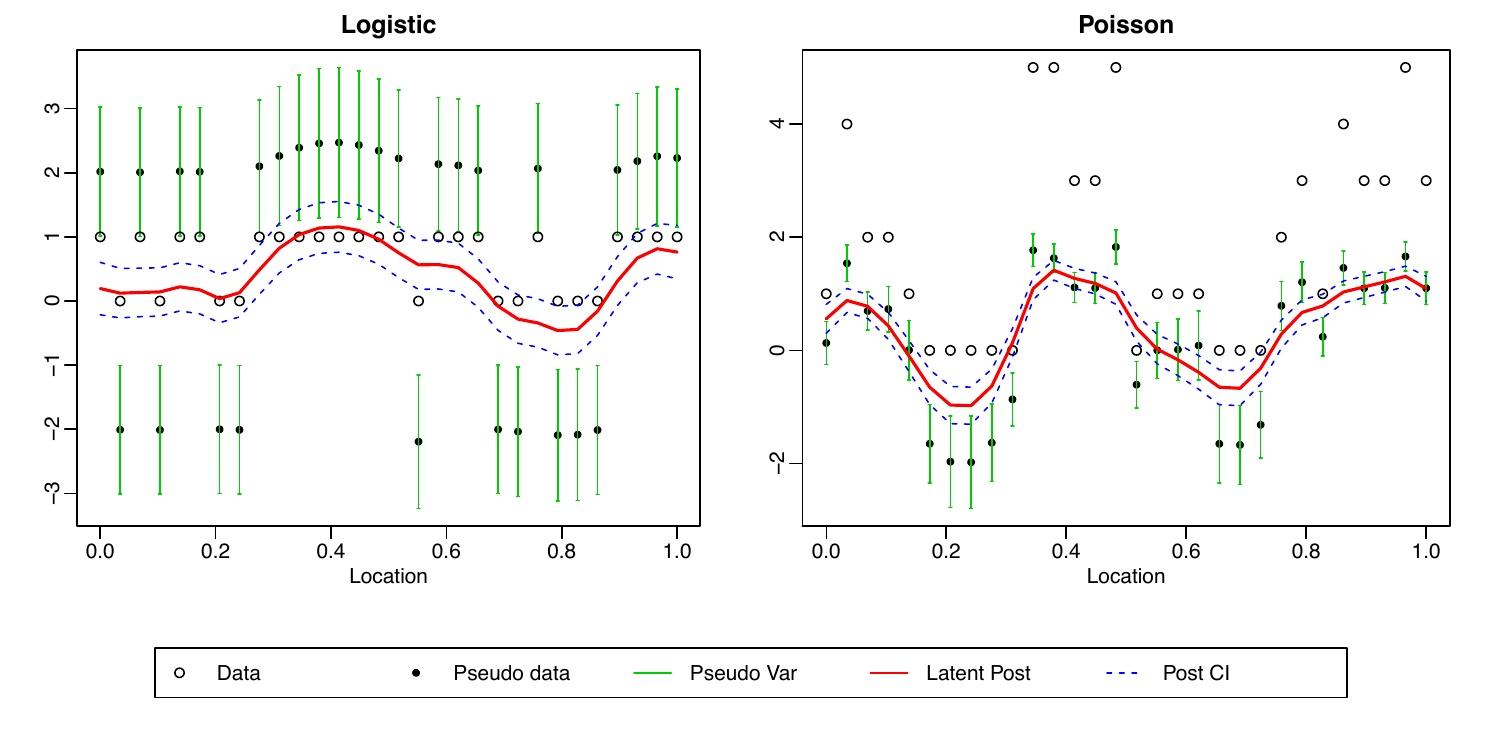}
  \caption{Pseudo-data $\bt_{\bfalpha}$ plus or minus half the standard deviation of the pseudo-noise for simulated data $\bz$ in one spatial dimension, along with the latent posterior mode $\bfalpha$ plus or minus half the posterior standard deviation. Note that the data exhibit a different scale than the pseudo-data due to the link function.}
\label{fig:pseudoillus}
\end{figure}

Some examples of popular likelihoods and the corresponding pseudo-data and pseudo-variances are summarized in Table \ref{tab:likelihoods}. The Bernoulli and Poisson cases are also illustrated in Figure \ref{fig:pseudoillus}.

Once the algorithm has converged (i.e., $ \bfalpha:=\by^{(\ell+1)} =\by^{(\ell)}$), we can use the second-order expansion of the loglikelihood at the mode as a Gaussian approximation of the likelihood based on pseudo-data,
\begin{equation}
\label{eq:laplacedata}
\widehat p_L(\bz\vert \by) = p(\bt_{\bfalpha} \vert \by) = \normal_n(\bt_{\bfalpha} \vert \by, \bD_{\bfalpha}),
\end{equation}
or combine it with the Laplace approximation to get a Gaussian approximation of the posterior conditional on pseudo-data,
\begin{equation}
\label{eq:laplacepseudo}
\widehat p_L(\by \vert \bz) = p(\by \vert \bt_{\bfalpha}) = \normal_n(\by|\bfalpha,\bW^{-1}_{\bfalpha}).
\end{equation}
%Intuition for interpreting the pseudo-data is discussed in Section \ref{supp:pseudo}.
For conciseness, we henceforth refer to \eqref{eq:laplacepseudo} as the ``Laplace approximation," rather than the more precise ``combined Gaussian and Laplace approximation."

%%%%%
\subsection{Review of the general Vecchia approximation \label{sec:vecchia}}

The Laplace approximation described in Section \ref{sec:laplace} allows us to deal with non-Gaussian likelihoods, but it still requires decomposing the $n \times n$ matrix $\bK$ and thus scales as  $\order(n^3)$. To achieve computational feasibility even for data sizes $n$ in the tens of thousands or more, we also apply a general Vecchia approximation \citep{Katzfuss2017a}, which we will briefly review here. 

Assume that $\by \sim \normal_n(\bfmu,\bK)$ is a vector of GP realizations and $\bt | \by \sim \normal_n(\by,\bD)$ a vector of noisy data, where $\bD$ is diagonal. 
Then, consider a vector $\bx = \by \cup \bt$ consisting of the $2n$ elements of $\by$ and $\bt$ in some ordering (more details below). 
It is well known that the density function, $p(\bx)$, can be factored into a product of univariate conditional densities,
$
p(\bx) = \prod_{i=1}^{2n} p(x_i|\bx_{1:i-1}).
$
The general Vecchia framework extends the approximation of \citet{Vecchia1988} to the vector $\bx$ consisting of latent GP realizations and noisy data, resulting in the approximate density 
\begin{equation}
\label{eq:vecchia}
\textstyle\widehat p(\bx) = \prod_{i=1}^{2n} p(x_i|\bx_{c(i)}),
\end{equation}
where $c(i) \subset \{1,\ldots,i-1 \}$ is a conditioning index set of size $m$ (or of size $i-1$ for $i \leq m$).
A small $m$ can lead to enormous computational savings and good approximations; \citet{Schafer2020} show that under some settings, the approximation error can be bounded when $m$ increases only polylogarithmically with $n$. Connections between Vecchia and composite likelihood \citep[e.g.,][]{varin2011overview} are discussed in \citet[][Sect.~3.8]{Katzfuss2017a}.

As $y_i = y(\bs_i)$ is indexed by location and $t_i$ is the corresponding noisy observation, the ordering within $\by$ and within $\bt$ is determined by an ordering of the observed locations, $\bs_1,\ldots,\bs_n$. We will use a coordinate-based (left-to-right) ordering in one spatial dimension.  In higher-dimensional spaces, we recommend a maxmin ordering \citep{Guinness2016a,Schafer2017}, which sequentially chooses each location in the ordering to maximize the minimum distance to previous locations in the ordering.

By straightforward extension of the proof of Prop.~1 in \citet{Katzfuss2017a} to the case $\bfmu \neq \bfzero$, it can be shown that the approximation in \eqref{eq:vecchia} implies a multivariate normal joint distribution, $\adens(\bx) = \normal(\bfmu_x,\bQ^{-1})$, where $\bfmu_{x,i} = \mu(\bs_j)$ if $x_i = y_j$ or $x_i = t_j$, $\bQ = \bU\bU'$, and $\bU$ is the sparse upper triangular Cholesky factor based on a reverse row-column ordering of $\bQ$. We write this as $\bU = \rchol(\bQ) \colonequals \rev(\chol(\rev(\bQ)))$, where $\rev(\cdot)$ reverse-orders the rows and columns of its matrix argument. The nonzero entries of $\bU$ are computed directly based on the covariance function $K$ as described in Appendix \ref{app:computeU}. 

Let $\bU_y$ and $\bU_t$ be the submatrices of $\bU$ consisting of the rows of $\bU$ corresponding to $\by$ and $\bt$, respectively. Then, the sparse matrix $\bW = \bU_y\bU_y'$ is the general Vecchia approximation to the posterior precision matrix of $\by$ given $\bt$. Defining $\bV \colonequals \rchol(\bW)$, we can obtain the posterior mean of $\by$ as $\E(\by|\bt) = \bfmu-(\bV')^{-1}\bV^{-1}\bU_y \bU_t'(\bt - \bfmu)$.

%%%%%%%%%%%%%%
\subsection{Ordering in Vecchia approximations\label{sec:specificvecchia}}

We now describe two specific approximations within the general Vecchia framework, which are based on how the elements of $\by$ and $\bt$ are ordered in the vector $\bx$ in \eqref{eq:vecchia}: Interweaved (IW) ordering and response-first (RF) ordering. While other ordering and conditioning schemes can also be used in the Vecchia-Laplace methodology to be introduced in Section \ref{sec:vlapprox}, we recommend these specific schemes to achieve high accuracy while ensuring linear complexity.

%%%
\subsubsection{Interweaved (IW) ordering \label{sec:iw}}

Vecchia-Interweaved (IW) is the sparse general Vecchia approach proposed for likelihood inference in \citet{Katzfuss2017a}, reviewed briefly here. It is a special case of general Vecchia in \eqref{eq:vecchia}, in which $\bx = (y_1,t_1,y_2,t_2,\ldots,y_n,t_n)'$ is specified using an interweaved ordering of the latent $\by$ and responses $\bt$. We consider the following specific expression for \eqref{eq:vecchia}:
\begin{equation}
\label{eq:vecchiaapprox}
\adens_{IW}(\bx)  = \prod_{i=1}^{n} p(t_i | y_{i} ) \, p(y_i | \by_{q_y(i)}, \bt_{q_t(i)}).
\end{equation}
If $x_j = t_i$, we only condition on $y_i$, because $\bD$ is diagonal and therefore $t_i$ is conditionally independent of all other variables in $\by$ and $\bt$ given $y_i$. If $x_j = y_i$, we condition on $\by_{q_y(i)}$ and $\bt_{q_t(i)}$, where $q(i) = q_y(i) \cup q_t(i)$ is the conditioning index vector consisting of the indices of the nearest $m$ locations previous to $i$ in the ordering. 
For splitting $q(i)$ into $q_y(i)$ and $q_t(i)$, we attempt to maximize $q_y(i)$ while ensuring linear complexity \citep{Katzfuss2017a}. %Specifically, for $i=1,\ldots,n$, we obtain $h(i) = \argmax_{j \in q(i)} |q_y(j) \cap q(i)|$, $k_i=\argmin_{\ell \in h(i)} \|\bs_i - \bs_\ell\|$, and then set $q_y(i) = (k_i) \cup (q_y(k_i) \cap q(i))$. 
%In words: among the nearest neighbors to the $i$th point, this procedure finds a neighbor with the most similar conditioning set.  Within that neighbors conditioning set, the element closest to point $i$ is tapped for its conditioning set, for which the intersection with the nearest neighbors determines the latent conditioning for point $i$.
Specifically, for $i=1,\ldots,n$, we set $q_y(i) = (k_i) \cup (q_y(k_i) \cap q(i))$, where $k_i \in q(i)$ is the index whose latent-conditioning set has the most overlap with $q(i)$: $k_i=\arg\max_{j \in q(i)} |q_y(j) \cap q(i)|$, choosing the closest $k_i$ in space to $\bs_i$ in case of a tie.
In one-dimensional space with coordinate ordering, this results in $q_y(i)= q(i) = (\max(1,i-m),\ldots,i-1)$ and $q_z(i) = \emptyset$. %This special case is referred to as latent-first with autoregressive conditioning (LF-auto) in \citet{Katzfuss2018}, where it is shown to have superlative performance for prediction in the 1D case due to a screening effect.  
In higher-dimensional space, we may not be able to condition entirely on $\by$, so the remaining conditioning indices are assigned to $q_t(i) = q(i) \setminus q_y(i)$. 
These conditioning rules guarantee that $\bU$ and $\bV$ are both highly sparse with at most $m$ nonzero off-diagonal elements per column. \citet{Katzfuss2017a} showed that these matrices, and the resulting posterior mean and precision matrix, can be obtained in $\order(nm^3)$ time.

%%%
\subsubsection{Response-first (RF) ordering\label{sec:dl}}

For approximating predictions at observed locations in Algorithm \ref{alg:vl} in more than one dimension, we recommend the new RF-full method described in \citet{Katzfuss2018}, reviewed briefly here. RF-full orders first all response variables, then all latent variables: $\bx = (\bt',\by')' = (t_1,\ldots,t_n,y_1,\ldots,y_n)'$. We consider the following specific expression for \eqref{eq:vecchia}:
\[
\adens_{RF}(\bx) = \prod_{i=1}^n \dens(t_i) \, \dens(y_i\vert \by_{q_y(i)}, \bt_{q_t(i)}).
\]
The responses $t_i$ do not condition on anything and are considered independent; this implies a poor approximation to $\dens(\bt)$, but it does not affect the posterior distribution $p(\by|\bt)$, which is the relevant quantity for our purposes. 
We now assume $q(i) = q_y(i) \cup q_t(i)$ to be set of indices corresponding to the $m$ locations closest to $\bs_i$ (including $\bs_i$), not considering the ordering. For any $j \in q(i)$, we then let $y_i$ condition on $y_j$ if it is ordered previously in $\bx$; otherwise, we condition on $t_j$. More precisely, we set $q_y(i) = \{j \in q(i): j<i\}$ and $q_t(i) = \{j \in q(i): j\geq i\}$.  Similar to IW, RF-full inference can be carried out in $\order(nm^3)$ time \citep{Katzfuss2018}. 
% using additional trick for $\bV$

%%%%%%%%%%%%%%%%%%%%%%%%%%%%%%%%%%%
\section{Vecchia-Laplace methods \label{sec:vlapprox}}

We now introduce our Vecchia-Laplace (VL) approximation, which allows fast inference for large datasets modeled using GGPs, by combining the Laplace and general Vecchia approximations reviewed in Section \ref{sec:review}.

%%%%
\subsection{The VL algorithm}

% At a high level, we take an initial guess of the posterior mode and begin Newton's method to maximize the log posterior density.  The posterior is written as a sum of the log likelihood and log prior; we approximate the exponential family likelihood with a second order Taylor expansion around our current guess for the posterior mode and approximate the joint prior and likelihood using the general Vecchia approximation.  The Newton step is simply the solution to a quadratic system and is relatively inexpensive to compute due to the Vecchia approximation.

To apply a Laplace approximation, it is first necessary to find the mode of the posterior density of $\by$. Rapid convergence to the mode can be achieved using a Newton-Raphson algorithm, which can be viewed as iteratively computing a new value $\by^{(l+1)}$ as the posterior mean of the latent GP realization $\by$ based on Gaussian pseudo-data $\bt = \bt_{\by^{(l)}}$, as discussed in Section \ref{sec:laplace}. Our VL algorithm applies a general Vecchia approximation $\widehat p(\bx)$ to the joint distribution of $\bx = \by \cup \bt$ at each iteration $l$, and computes the posterior mean of $\by$ given $\bt$ under this approximate distribution. We recommend IW ordering (Section \ref{sec:iw}) in one spatial dimension, and RF ordering (Section \ref{sec:dl}) when working in more than one dimension.  The resulting VL algorithm is presented as Algorithm \ref{alg:vl}. After convergence, we obtain the approximation
\begin{equation}
\label{eq:vecchialaplace}
\widehat p_{VL}(\by|\bz) = \normal_n(\by|\bfalpha_V, \bW_V^{-1}).
\end{equation}

\begin{algorithm}
 \caption{Vecchia-Laplace (VL)}
\begin{algorithmic}[1]
\Procedure{Vecchia-Specify}{$\locs, m$} \Comment{Define Vecchia Structure}
	\State Order locations $\locs$ using coordinate (in 1D) or maxmin ordering (in 2D or higher)\;
	\State For VL-IW, determine variable ordering and conditioning as in Sect.~\ref{sec:iw}\;
	\State For VL-RF, determine variable ordering and conditioning as in	Sect.~\ref{sec:dl}\;
	\State \Return ordering and conditioning info in Vecchia Approximation Object $\mathtt{VAO}$\;
\EndProcedure
\Statex
\Procedure{VL-Inference}{$\bz, \mathtt{VAO}, g_i, \bfmu, K$} \Comment{Maximize GP Posterior}	
	\State Derive $u_i(\cdot) =  \frac{\partial}{\partial y} \log g_i \big|_{(\cdot)}$ and $d_i(\cdot) =-\big(\frac{\partial^2}{\partial y^2} \log g_i\big)^{-1} \big|_{(\cdot)}$
	\State Initialize $\by^{(0)} = \bfmu $ \;
	\For{$l$=0,1,\ldots}
		\State Compute $\bu = (u_1(y_1^{(l)}),\ldots,u_n(y_n^{(l)}))'$ and $\bD = \diag(d_1(y_1^{(l)}),\ldots,d_n (y_n^{(l)}))$ \;
		\State Update pseudo-data $\bt = \by^{(l)} + \bD \bu$ \;
		\State Compute $\bU$ (see Appendix \ref{app:computeU}) based on $\bD$, $K$, and $\mathtt{VAO}$  \;
		\State Extract submatrices $\bU_y$ and $\bU_t$ \;
		\State Compute $\bW = \bU_y\bU_y'$ and $\bV = \rchol(\bW)$ \;
		\State  Compute the new posterior mean: $\by^{(l+1)} = \bfmu - (\bV')^{-1}\bV^{-1}\bU_y \bU_t'(\bt - \bfmu)$ \;
 		\If{ $\|\by^{(l+1)}-\by^{(l)}\| < \epsilon$ }\;
			\State \Return $\bfalpha_V = \by^{(l+1)}$ and $\bW_V = \bW$ \; \Comment{Posterior Mode Estimate}
		\EndIf
	\EndFor
\EndProcedure
\end{algorithmic}
\label{alg:vl}
\end{algorithm}

Once the algorithm has converged and the posterior mean $\bfalpha_V$ and precision $\bW_V$ have been obtained, the posterior distribution in \eqref{eq:vecchialaplace} can be used for estimation of the integrated likelihood (Section \ref{sec:likelihood}) and for prediction at unobserved locations (Section \ref{sec:prediction}). As we will see in our simulation studies later, even for moderate $m$, the VL procedure in Algorithm \ref{alg:vl} essentially finds the exact mode of the posterior.

%%%%%%%
\subsection{Integrated likelihood for parameter inference \label{sec:likelihood}}

In the case of unknown parameters $\bftheta$ in $\mu$, $K$, or in the $g_i$, we would like to carry out parameter inference based on the integrated likelihood,
\[
\mathcal{L}(\bftheta) = p(\bz|\bftheta) = \int p(\bz|\by,\bftheta) p(\by|\bftheta) d\by.
\]
However, this quantity is exactly the unknown normalizing constant in the denominator of \eqref{eq:exactposterior}, and the integral can generally not be carried out analytically. Instead, we will base parameter inference on the integrated likelihood implied by our VL approximation.  In the following, we will again suppress dependence on $\bftheta$ for ease of notation.  

First, rearranging terms in \eqref{eq:exactposterior}, we have $p(\bz) = p(\bz|\by) p(\by)/p(\by|\bz)$. The Laplace approximation approximates the posterior in the denominator as $\adens_L(\by|\bz) = p(\by|\bt_{\bfalpha})$ (see \eqref{eq:laplacepseudo}). Noting that rearranging the definition of a conditional density gives $p(\by) = p(\by,\bt)/p(\bt|\by)$, we obtain the Laplace approximation of the integrated likelihood:
\begin{equation}
\label{eq:laplacelikelihood}
\mathcal{L}_L(\bftheta) = \adens_L(\bz) = \frac{p(\by,\bt)}{ p(\by \vert \bt)} \cdot \frac{p(\bz \vert \by)}{p(\bt \vert \by)} = p(\bt) \cdot \frac{p(\bz \vert \by)}{p(\bt \vert \by)} ,
\end{equation}
where the terms are evaluated at $\by = \bfalpha$ and $\bt = \bt_{\bfalpha}$. In this form, the approximation of the integrated likelihood of the data $\bz$ can be interpreted as a product of the integrated likelihood of the Gaussian pseudo-data $p(\bt)$, times a correction term given by the ratio of the true likelihood to the Gaussian likelihood of the pseudo-data: $p(\bz \vert \by)/p(\bt \vert \by) = \prod_{i=1}^n g_i(z_i|y_i)/\normal(t_i|y_i,d_i)$.

To achieve scalability, we approximate the density $p(\bt) = p(\bx)/p(\by \vert \bt)$ as implied by the IW approximation $\adens_{IW}(\bx)$ in \eqref{eq:vecchiaapprox}. The resulting expression for $\adens_{IW}(\bt)$ is derived in \citet{Katzfuss2017a} for the case of $\bfmu =\bfzero$.  We show in Section \ref{supp:vllikelihood} that the approximate density essentially has the same form if the prior mean is not zero: 
\[
\textstyle -2 \log \adens_{IW}(\bt) = -2\sum_i \log \bU_{ii}+2\sum_i \log \bV_{ii}+ \tilde\bt'\tilde\bt - \breve\bt'\breve\bt + n\log(2\pi),
\]
where $\tilde\bt = \bU_t'(\bt - \bfmu)$ and $\breve\bt = \bV^{-1}\bU_y \tilde\bt$.

Thus, for a specific parameter value $\bftheta$, we run Algorithm \ref{alg:vl} based on $\bftheta$ to obtain $\bfalpha_V$, set $\by= \bfalpha_V$, $\bt = \bt_{\bfalpha_V}$, and $d_i = (\bD_{\bfalpha_V})_{ii}$, and then evaluate the VL integrated likelihood as
\begin{equation}
\label{eq:vllikelihood}
\mathcal{L}_{VL}(\bftheta) = \adens_{VL}(\bz|\bftheta) = \adens_{IW}(\bt) \,\, \prod_{i=1}^n \frac{g_i(z_i|y_i)}{\normal(t_i|y_i,d_i)}.
\end{equation}
We can plug $\mathcal{L}_{VL}(\bftheta)$ into any numerical likelihood-based inference procedure, such as numerical optimization for finding the maximum likelihood estimator of $\bftheta$, or sampling-based algorithms for finding the posterior of $\bftheta$. In an iterative inference procedure, we recommend initializing $\by^{(0)}$ in Algorithm \ref{alg:vl} at the mode $\bfalpha_V$ obtained for the previous parameter value.
Our integrated likelihood can also be used directly to evaluate the posterior of $\bftheta$ over a grid of high-probability points \citep[][Sect.~3.1]{Rue2009}. An extension to the integrated nested Laplace approximation (INLA) that improves the accuracy of the marginal posteriors of the $y_i$ \citep[][Sect.~3.2]{Rue2009} is straightforward.

%%%%%%%%%%
\subsection{Predictions at unobserved locations \label{sec:prediction}}

We now consider making predictions at $n^\star$ unobserved locations, $\locs^\star = \{\bs_1^*,\ldots,\bs_{n^\star}^*\}$, by obtaining the posterior distribution of $\by^\star = (y_1^\star,\ldots,y_{n^\star}^\star)'$ with $y_i^\star=y(\bs_i^\star)$.
Using the Laplace approximation as expressed in \eqref{eq:laplacedata}, GGP predictions are approximated as GP predictions given Gaussian pseudo-data $\bt_{\bfalpha}$ with noise covariance matrix $\bD_{\bfalpha}$. 

Hence, to obtain scalable predictions at unobserved locations, we use the recommended prediction methods in \citet{Katzfuss2018} that apply Vecchia approximations to the multivariate normal vector $\widetilde\bx = \bt \cup \by \cup \by^\star$. For one-dimensional space, we use an extension of IW called LF-auto in \citet{Katzfuss2018}, and for higher-dimensional space we use the RF-full method of \citet{Katzfuss2018}. In both cases, the pseudo-data $\bt = \bt_{\bfalpha_V}$ and the noise variances $\bD = \bD_{\bfalpha_V}$ are evaluated at the approximate mode $\bfalpha_V$ obtained using Algorithm \ref{alg:vl}. 
% Defining $\widetilde\by = (\widetilde y_1,\ldots,\widetilde y_{n+n^\star})' = (y_1,\ldots,y_n,y_1^\star,\ldots,y_{n^\star}^\star)'$, the VL-RF approximation takes the form
% \[
% \textstyle \adens_{RF}(\bx)  = \big(\prod_{i=1}^{n} p(t_i ) \big) \big(\prod_{i=1}^{n+n^\star} p(\widetilde y_i | \widetilde\by_{q_y(i)}, \bt_{q_t(i)}) \big),
% \]
% as described in Section \ref{sec:dl}, except that for $i>n$, we set $q(i)$ to consist of the indices corresponding to the $m$ nearest locations to $\bs_i$ \emph{excluding} $\bs_i$.
Based on this approximation, we can compute the implied posterior distribution of $\widetilde\by = \by \cup \by^\star$ as described in Section \ref{sec:vecchia}: $\adens(\widetilde\by | \bt) \sim \normal(\widetilde\bfmu,(\widetilde\bV\widetilde\bV')^{-1} )$. \citet{Katzfuss2018} describe how to efficiently extract quantities of interest from this distribution, including the posterior mean and variances at unobserved locations.
Finally, summaries or samples from the posterior of $\widetilde\by$ can be transformed to the data scale using the likelihood function $g(z|y)$, if desired. Sometimes it is difficult to compute certain predictive summaries at the data scale analytically, but it is always possible to approximate them via sampling.

Algorithm \ref{alg:practical} in Section \ref{app:algo} provides pseudo-code for maximum-likelihood estimation of parameters and for prediction.

%%%%%%%%%
\subsection{Properties \label{sec:propts}}

\subsubsection{Complexity}

Inference for GPs with independent Gaussian noise using the Vecchia approximations considered here requires $\order(nm^3)$ time, where $m$ is the maximum size of the conditioning sets $q(i)$, and can be easily parallelized \citep{Katzfuss2017a,Katzfuss2018}. Our VL Algorithm \ref{alg:vl} iteratively computes the Vecchia approximation multiple times until convergence, only adding $\order(n)$ cost at each iteration for computing the pseudo-data $\bt_{\by^{(l)}}$. Hence, the VL algorithm requires $\order(knm^3)$ time, where $k$, the number of iterations required until convergence, can be very small (often, $k<10$).

Once $\bfalpha_V$ has been determined using Algorithm \ref{alg:vl}, evaluating the integrated likelihood \eqref{eq:vllikelihood} for parameter inference requires $\order(nm^3)$ time \citep{Katzfuss2017a}, and prediction at $n^\star$ unobserved locations requires $\order((n+n^\star)m^3)$ time \citep{Katzfuss2018}. Thus, all computational costs are linear in $n$ for fixed $m$.

%%%%%
\subsubsection{Approximation errors\label{sec:error}}

Our VL approximation $\adens_{VL}(\by|\bz)= \normal_n(\by|\bfalpha_V, \bW_V^{-1})$ in \eqref{eq:vecchialaplace} has two sources of error relative to the true posterior $p(\by|\bz)$: the Vecchia approximation and the Laplace approximation. Both errors are difficult to quantify in general, but our numerical experiments in Section \ref{sec:simulation} show that our approximation can be very accurate.
The error due to the Vecchia approximation can always be reduced by increasing $m$ \citep[e.g.,][]{Katzfuss2018}. 
% In the special case of Gaussian observation distributions, the VL algorithm will converge in a single iteration, as the pseudo-data and -variances do not depend on $\by$ and are equal to the actual data and noise variances, respectively (see Table \ref{tab:likelihoods}). In addition, the data likelihoods in the integrated likelihood in \eqref{eq:laplacelikelihood} will cancel.
%and the results will be identical to the general Vecchia likelihood \citep{Katzfuss2017a} and general Vecchia prediction \citep{Katzfuss2018}.

The error of the Laplace approximation is known to depend on the likelihood being approximated. Laplace is exact for Gaussian likelihoods, in which case the VL approximation reverts to the general Vecchia approximation. For non-Gaussian spatial data, theoretical error bounds are difficult to obtain \citep[e.g.,][Sect.~4.1]{Rue2009}. From an empirical point of view, \citet{fong2010bayesian} affirm the non-spatial results of \citet{breslow1993approximate}, showing that INLA, an extension of the Laplace approximation, generally performs well for GGPs, with the exception of some types of binomial data. \citet{bonat2016practical} provide a thorough simulation study comparing Laplace to MCMC methods for parameter estimation in the case of binomial, Poisson, and negative-binomial spatial data; they conclude that the Laplace approximation is ``a safe option" that is computationally practical.

\subsubsection{Convergence}

For GGPs as described in Section \ref{sec:GGP}, the log-posterior in \eqref{eq:exactposterior} is concave under appropriate parameterizations. Existing results show that the Newton-Raphson algorithm used in the Laplace approximation is then theoretically guaranteed to converge to its mode \citep[e.g.,][Section 9.5.2]{boyd2004convex}. In our VL Algorithm \ref{alg:vl}, the distribution $\adens(\by)$ implied by the general Vecchia approximation changes at each iteration, which makes it difficult to theoretically guarantee convergence, except in special cases.
% Unfortunately, it is difficult to prove that Algorithm \ref{alg:vl} is guaranteed to converge, except in special cases. As discussed in Section \ref{sec:approxerrors}, the VL updates at each iteration of Algorithm \ref{alg:vl} are exact for $m\geq n-1$, and virtually exact for sufficiently large $m$, so in those case convergence is (virtually) guaranteed. The case $m=0$ is equivalent to assuming that the $y_i$ are independent, which also leads to a concave log-posterior and hence guaranteed convergence. However, for small $m>0$ in between these special cases, it is difficult to theoretically guarantee convergence or even to determine what objective function is being optimized, because the approximate distribution $\adens(\by)$ depends on pseudo-variances $\bD_{\by^{(l)}}$ that change at each iteration $l$, except in the Gaussian case.
Fortunately, empirical testing of Algorithm \ref{alg:vl} under different parameter and data settings showed that convergence can always be expected when machine precision is not an issue.

%%%%%%%%%%%%%%%%%%%%%%%%%%%%%%%%%%%%%%%%%%%%%%%%%%%%%%%
\section{Simulations and comparisons \label{sec:simulation}}

We compared our VL approaches to other methods using simulated data. Throughout Section \ref{sec:simulation}, unless specified otherwise, we simulated realizations $\by$ on a grid of size $\sqrt{n}\times \sqrt{n}$ on the unit square from a GP with mean zero and a Mat\'ern covariance function with variance 1, smoothness $\nu$, and range parameter $\lambda = 0.05$.  Gridded locations allow us to carry out simulations for large $n$ using Fourier methods. The data were then generated conditional on $\by$ using the four likelihoods in Table \ref{tab:likelihoods}, with $a=2$ in the Gamma case.

As low-rank approximations are very popular for large spatial data, we also considered a fully independent conditional or modified-predictive-process approximation to Laplace with $m$ knots (abbreviated as LowRank here), which is equivalent to VL-IW except that each conditioning set $q_y(i) = (1,\ldots,m)$ simply consists of the first $m$ latent variables in maxmin ordering. This equivalence allowed us to run VL and LowRank using the same code base, thus avoiding differences solely due to programming.

Criteria used for comparison are the run time (on a 2017 MacBook Pro), the relative root mean square error (RRMSE) and the difference in log scores (dLS). Results are averaged over 100 simulated datasets, unless noted otherwise. The RRMSE is the root mean square error of the posterior mean of $\by$ obtained by one of the approximation methods relative to the true simulated $\by$, divided by the RMSE of the Laplace approximation. The log score is computed as the negative logarithm of the approximated posterior density of $\by$ evaluated at the true $\by$, with low values corresponding to well calibrated and sharp posterior distributions \citep[e.g.,][Sect.~3]{gneiting2014probabilistic}. The dLS is the log score of an approximation method minus the log score for the Laplace approximation. When averaged over a sufficient number of simulated data, the dLS can be shown to approximate the difference between the Kullback-Leibler (KL) divergence of the exact posterior distribution and the considered approximation, minus the KL divergence between the exact distribution and the Laplace approximation.
%$$\text{MSE} = \frac{1}{n} \sum_i^n (\alpha_i - y_i)^2 $$
%$$\text{LS} =\log(p(\by|\bfalpha,  \bW_\alpha^{-1})) \sim \log \normal(\by | \bfalpha, \bW_\alpha^{-1}) $$
%where $\bfalpha = (\alpha_1, ..., \alpha_n )$ is the posterior mean and $\bfalpha$ and $\by = (y_1,...,y_n)$ is the true latent prior realization. 

%%%%%%%%
\subsection{Comparison to MCMC \label{sim:HMC}}

\begin{figure}
\centering
\includegraphics[scale = .75]{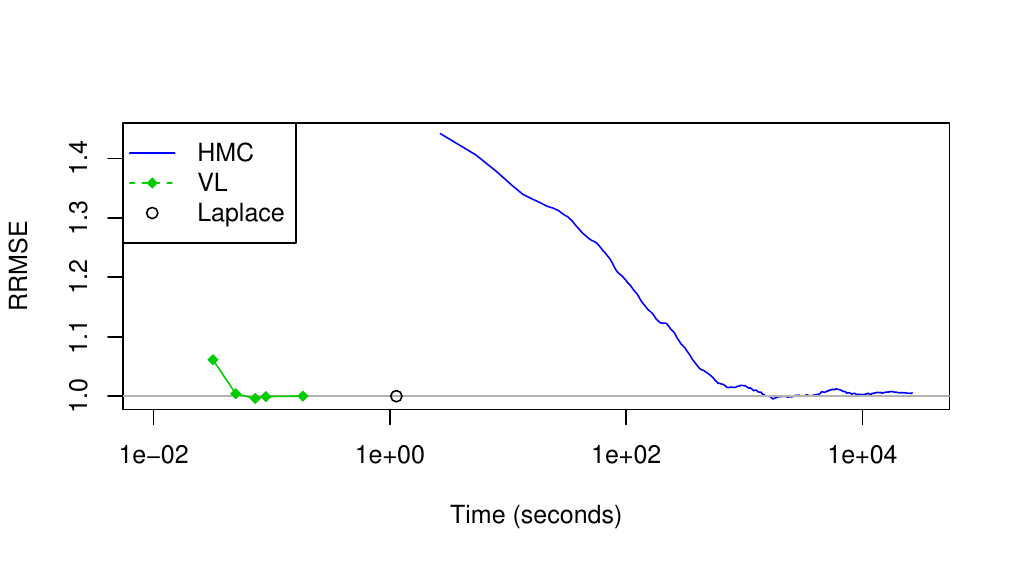}
 \caption{RRMSE versus time (on a log scale) for Bernoulli data of size $n=625$ on the unit square. Laplace is run once until convergence. For VL-RF, we considered $m \in \{ 1,5,10,20,40\}$. The number of HMC iterations varies from 10{,}100 to 1{,}000{,}000 in increments of 100, with the first 10{,}000 considered burn-in.}
 \label{fig:HMC}
\end{figure}

Non-Gaussian spatial models are often fitted using Markov chain Monte Carlo (MCMC), which under mild regularity conditions is ``exact approximate,'' converging to the true posterior as the number of iterations approaches infinity. For finite computation time and large $n$, however, MCMC results can be very poor relative to the Laplace approximation.
We demonstrate this with a single simulated dataset consisting of $n=625$ Bernoulli observations based on a GP with smoothness parameter $\nu=0.5$ on the unit square.  We compared Laplace and VL-RF to Hamiltonian Monte Carlo \citep[HMC;][]{neal2011mcmc}, a MCMC method well suited to sampling correlated variables. 
As shown in Figure \ref{fig:HMC}, VL quickly achieved the same accuracy as Laplace as $m$ increased, but at a fraction of the computing time. In contrast, HMC took orders of magnitude longer to achieve similar accuracy. Even with 1 million iterations, the RMSE for HMC was slightly higher than for VL; this is in line with existing simulation studies suggesting that the Laplace approximation error may be negligible in many GGP settings (see Section \ref{sec:error}).
We expect the relative performance of HMC to degrade further as $n$ increases. More details and results can be found in Section \ref{supp:sim}.

%%%%
\subsection{Computational scaling of Laplace approximations \label{sim:time}}

\begin{figure}[tb]
\centering
\includegraphics[trim=0mm 3mm 0mm 5mm, clip, width = .99\linewidth]{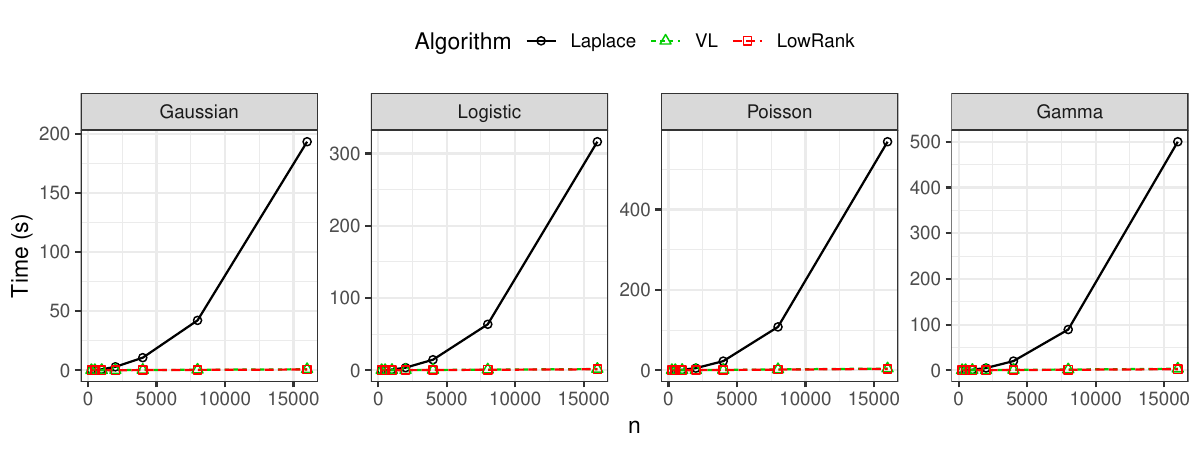} %lbrt
 \caption{For sample size $n$ between 250 and 16,000, computing time for the Laplace approximation based on Newton-Raphson, compared to VL and LowRank using Algorithm \ref{alg:vl} with $m=10$}
 \label{fig:time}
\end{figure}

While the Laplace approximation is very useful for moderate data sizes $n$, we now briefly illustrate the computational infeasibility for large $n$ due to its cubic scaling. In Figure \ref{fig:time}, we show the average computation time for observations with smoothness $\nu=0.5$ in the setting described later in Section \ref{sec:2dsim}. Clearly, Laplace using Newton-Raphson quickly became infeasibly slow as $n$ increased. In contrast, VL and LowRank scaled roughly linearly. 

%%%%
\subsection{VL accuracy in one-dimensional space \label{sec:1dsim}}

We now compare the accuracy of the VL and LowRank approximations. Both approaches scale linearly in $n$ for fixed $m$, and both approaches converge to the Laplace approximation as $m$ increases, with equivalence guaranteed for $m=n-1$.

\begin{figure}
\centering
	\begin{subfigure}{1\textwidth}
	\centering
	\includegraphics[width =.95\linewidth]{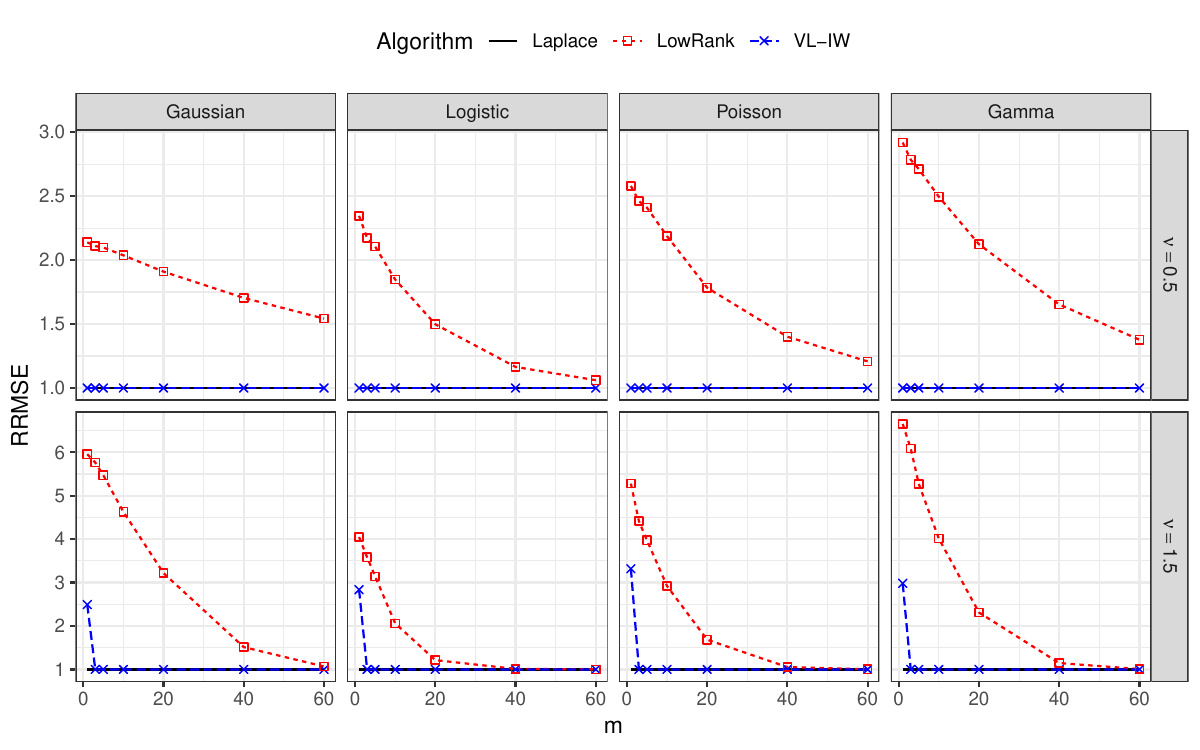}
	\caption{RMSE (relative to Laplace)}
	\end{subfigure}%

\vspace{3mm}

	\begin{subfigure}{1\textwidth}
	\centering
	\includegraphics[width =.95\linewidth]{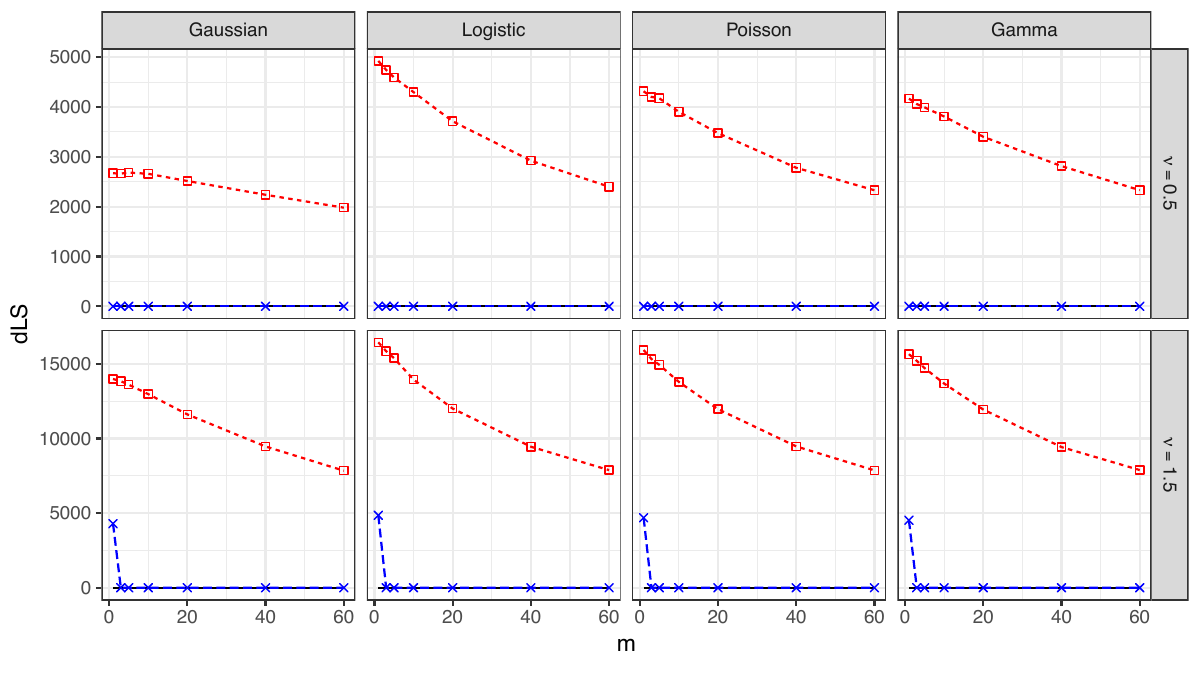}
	\caption{Difference in log score (relative to Laplace)}
	\end{subfigure}
 \caption{Simulation results for $n=2{,500}$ observations on a \textbf{one-dimensional} spatial domain}
 \label{fig:sim1d}
\end{figure}

% As explained in Section \ref{sec:vlapprox}, we use interweaved ordering for 1D analysis.  
Figure \ref{fig:sim1d} shows the average results for 100 simulated datasets of size $n=2{,}500$ each, on the unit interval. For the Gaussian likelihood, the noise variance was $\tau^2 = 0.1^2$.
Clearly, VL-IW was extremely accurate and delivered essentially equivalent results to the Laplace approximation, even for very small $m$. For exponential covariance (i.e., Mat\'ern with smoothness $\nu = 0.5$), an exact screening effect holds in one-dimensional space, and so VL-IW is exactly equal to Laplace for any $m \geq 1$. LowRank required much larger $m$ to achieve equivalence to Laplace.

%%%%
\subsection{VL accuracy in two-dimensional space \label{sec:2dsim}}

\begin{figure}
\centering
	\begin{subfigure}{1\textwidth}
	\centering
	\includegraphics[width =.95\linewidth]{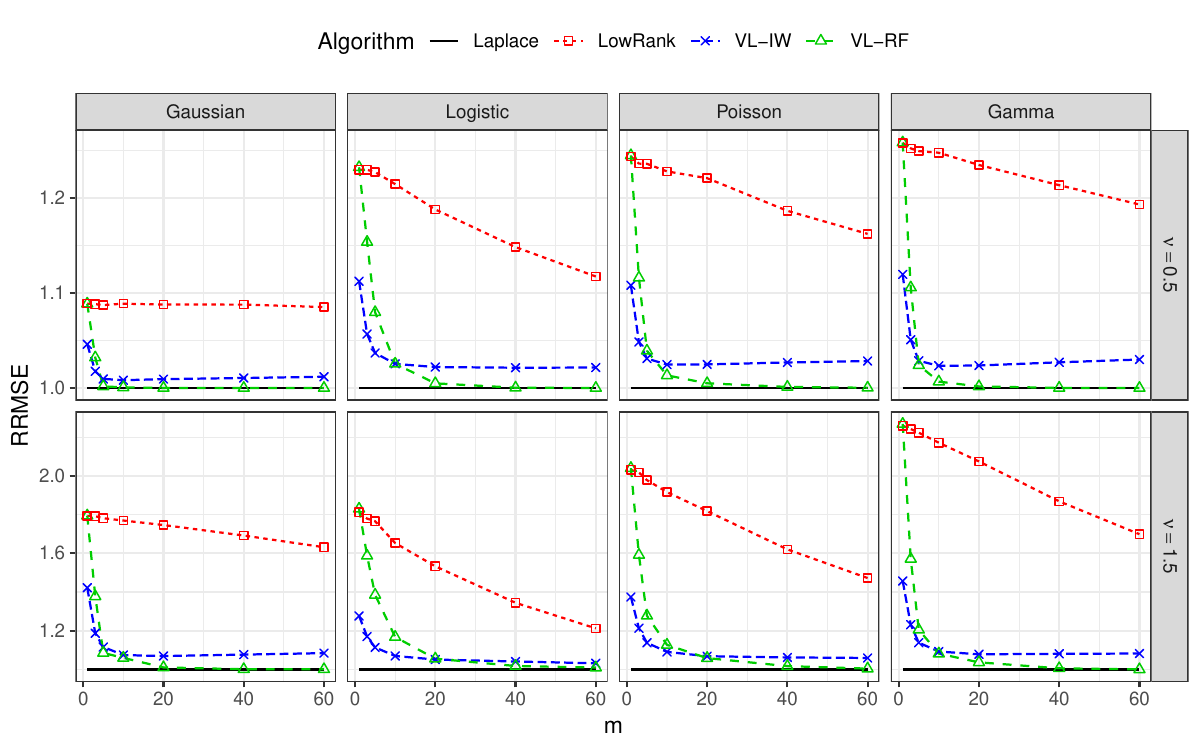}
	\caption{RMSE (relative to Laplace)}
	\end{subfigure}%

\vspace{3mm}

	\begin{subfigure}{1\textwidth}
	\centering
	\includegraphics[width =.95\linewidth]{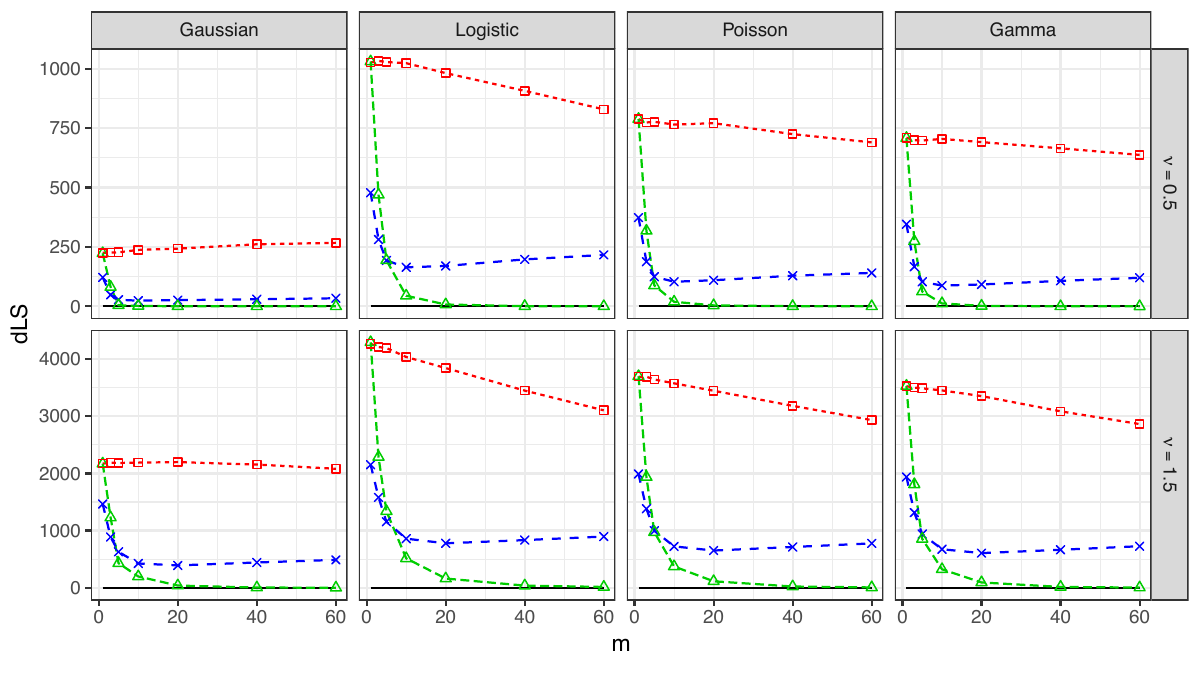}
	\caption{Difference in log score (relative to Laplace)}
	\end{subfigure}
 \caption{Simulation results for $n=2{,}500$ observations on a \textbf{two-dimensional} spatial domain}
 \label{fig:sim2d}
\end{figure}

Figure \ref{fig:sim2d} shows results for the same simulation study as in Section \ref{sec:1dsim}, except that the data were simulated on the two-dimensional unit square, with noise variance $\tau^2=0.1$ for the Gaussian likelihood.
%We used the obs-lat ordering as described in Section \ref{sec:vlapprox} but include the interweaved ordering when plotting the log score to show how obs-lat provides superior performance.  
While all methods are again equivalent to Laplace for $m=n-1$, the two-dimensional problem is considerably more difficult, and higher values of $m$ were required for accurate approximations. As we can see, the recommended VL-RF had roughly equivalent performance to Laplace once $m$ reached 20, and it was more accurate than VL-IW for $m>10$. LowRank performed considerably worse than the VL methods, and further simulations (not shown) showed that in some cases LowRank approached the accuracy of Laplace only when $m$ was almost as large as $n$.
A simulation with larger range parameter $\lambda = 0.2$ is shown in Section \ref{supp:sim2D_rangep2} of the supplement; while VL-RF was still more accurate than LowRank for all settings, the larger range reduced the amount of fine-scale variation, thus reducing the advantage of VL over LowRank relative to Figure \ref{fig:sim2d}, especially for logistic models. 
The relative performance of the methods was similar in higher dimensions; plots for 3 and 4 dimensions are shown in Section \ref{sec:simhigherdim}.

%We also demonstrate how a change in the domain or sample size does not require a change in the number of conditioning points, while a low rank approach demands additional knots and tuning as the domain and sample size grows or else the MSE degrades.  In this sense, VL has a simpler implementation (less tuning) for analyzing arbitrary datasets when compared to LL. 

\begin{figure}
\centering
	\includegraphics[trim=0mm 3mm 0mm 5mm, clip, width = .95\linewidth]{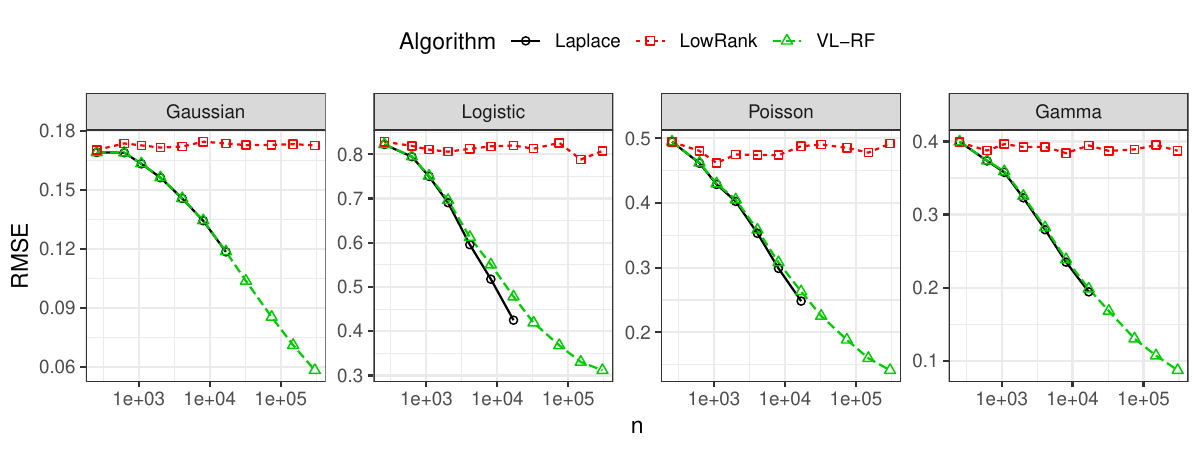}
\caption{On a two-dimensional domain with $\nu=0.5$ and fixed $m=10$, RMSE between true $\by$ and posterior mode $\bfalpha_V$ for increasing sample size $n$ (on a log scale) up to $300{,}000$. Laplace without further approximation becomes prohibitively expensive for large $n$, so we only computed it up to $n=16{,}000$.}
 \label{fig:mse_m}
\end{figure}

For larger $n$, the differences between LowRank and VL became even more pronounced. Figure \ref{fig:mse_m} shows the RMSE for simulations with increasing sample size $n$ but fixed $m$. VL-RF improved in accuracy under this asymptotic in-fill scenario almost as fast as Laplace, while LowRank failed to improve.

%%%%
\subsection{Simulations for log-Gaussian Cox processes\label{app:lgcp}}
   
Point patterns are sets of points or locations $\bs_1,\ldots,\bs_N$ in a domain $\domain$. 
%Examples include earthquake epicenters, trees, disease occurrences, or terror attacks. An observed point pattern can be thought of as a realization of a (stochastic) point process, which is a set of (consistent) probability distributions for random variables of the form $z(A)$, defined as the number of points in $A \subset \domain$.
%
%A point process is called a Poisson process with intensity function $\lambda(\cdot)$ if
%\begin{enumerate}
%\item $z(A) \sim Poisson(\mu(A))$ with $\mu(A) = \int_A \lambda(\bs) d\bs$ for any $A \subset \domain$, and
%\item $z(A_1),\ldots,z(A_k)$ independent for disjoint subsets $A_1,\ldots,A_k$.
%\end{enumerate}
A popular model for point patterns is the log-Gaussian Cox process (LGCP), a doubly stochastic Poisson process whose intensity function $\lambda(\cdot)$ is modeled as random, $\log \lambda(\cdot) = y(\cdot) \sim \GP(\mu,C)$. Inference for LGCPs is difficult due to stochastic integrals. 

A natural approximation \citep[e.g.,][]{Diggle2013} for LGCPs relies on partitioning the domain $\domain$ into $n$ grid cells $A_1,\ldots,A_n$ with center points $\ba_1,\ldots,\ba_n$, respectively.  The number of observed points falling into $A_i$ is treated as the data, $z_i = z(A_i) = \sum_{j=1}^N 1_{\bs_j \in A_i}$.  These gridded data conditionally follow a Poisson distribution,  $z_1,\ldots,z_n \,| \, y(\cdot)\, \stackrel{ind.}{\sim} \mathcal{P}(\mu(A_i))$, where 
\[
\textstyle \mu(A_i) = \int_{A_i} \lambda(\bs) d\bs \approx |A_i| \, \lambda(\ba_i) = |A_i| \, e^{y(\ba_i)}.
\]
%Smaller grid cells offer a more accurate approximation, so $n$ is often much larger than the number of observed points and many of the $z_i$ are zero.
This model falls under the GGP framework, so we can apply our VL methods to obtain fast inference for point patterns.

\begin{figure}[ht]
\centering
\includegraphics[width = .98\linewidth]{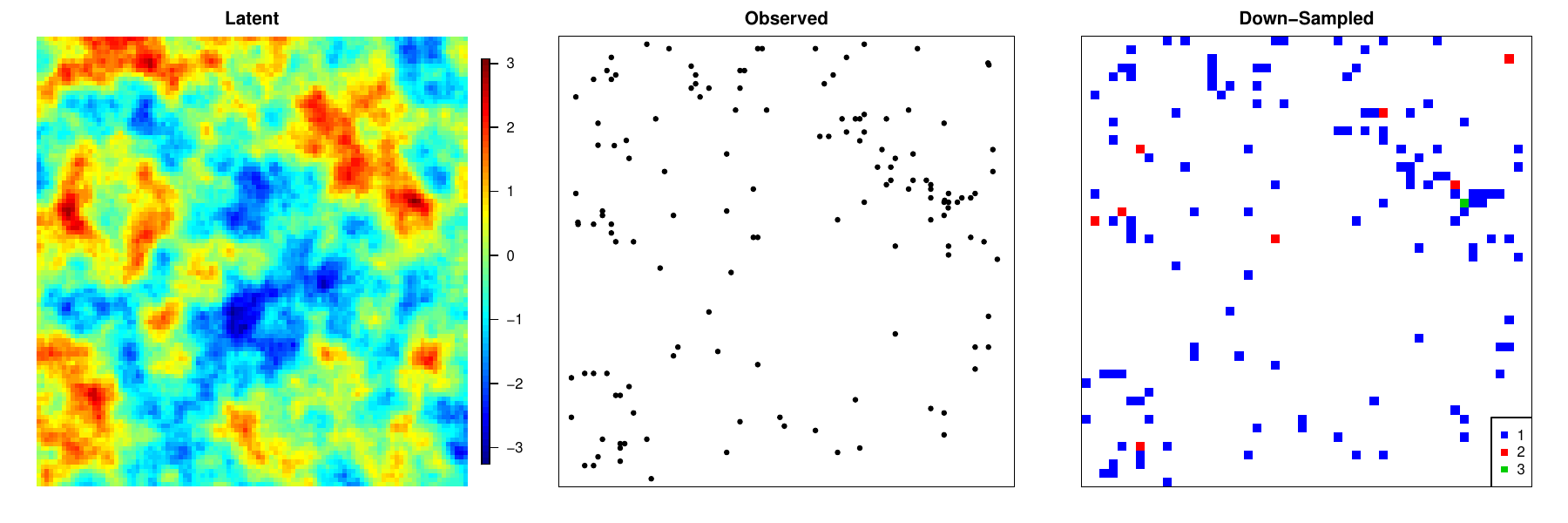}
	 \caption{Gridding a simulated LGCP point pattern: The latent log-intensity $y(\cdot)$ (left), a corresponding simulated point pattern (center), and the down-sampled Poisson count data used for analysis on a $n = 50 \times 50 = 2{,}500$ grid (right)}
\label{supp:LGCP_sim}
\end{figure}

Figure \ref{supp:LGCP_sim} shows a LGCP whose log-intensity is modeled as a GP with Mat\'ern covariance with range parameter 2.5 on a spatial domain $\domain = [0,50]^2$, discretized into $n = 2{,}500 = 50 \times 50$ unit-square grid cells. This is equivalent to the simulation in Section \ref{sec:2dsim}, as the domain can be scaled to a unit-square domain with range 0.05, but on the original scale the grid induces areal regions with unit area, $|A_i|=1$, with intensity function $\mu(A_i) = \exp(y(\ba_i))$. Thus, the averaged results for fitting repeatedly simulated datasets from this LGCP are equivalent to the Poisson results shown in the third column of Figure \ref{fig:sim2d}, indicating that VL can be used to obtain virtually equivalent inference to that using a Laplace algorithm, albeit at much lower computational cost for large $n$.

%%%%
\subsection{Parameter estimation \label{sim:parameters}}

We also explored parameter estimation based on each method's integrated-likelihood approximation. Specifically, we considered Poisson data at $n=625$ locations in the unit square, based on a GP with true smoothness $\nu = 0.5$ and range $\lambda= 0.05$. 

\begin{figure}
\centering
	\begin{subfigure}{1\textwidth}
	\centering
\includegraphics[width=.95\linewidth]{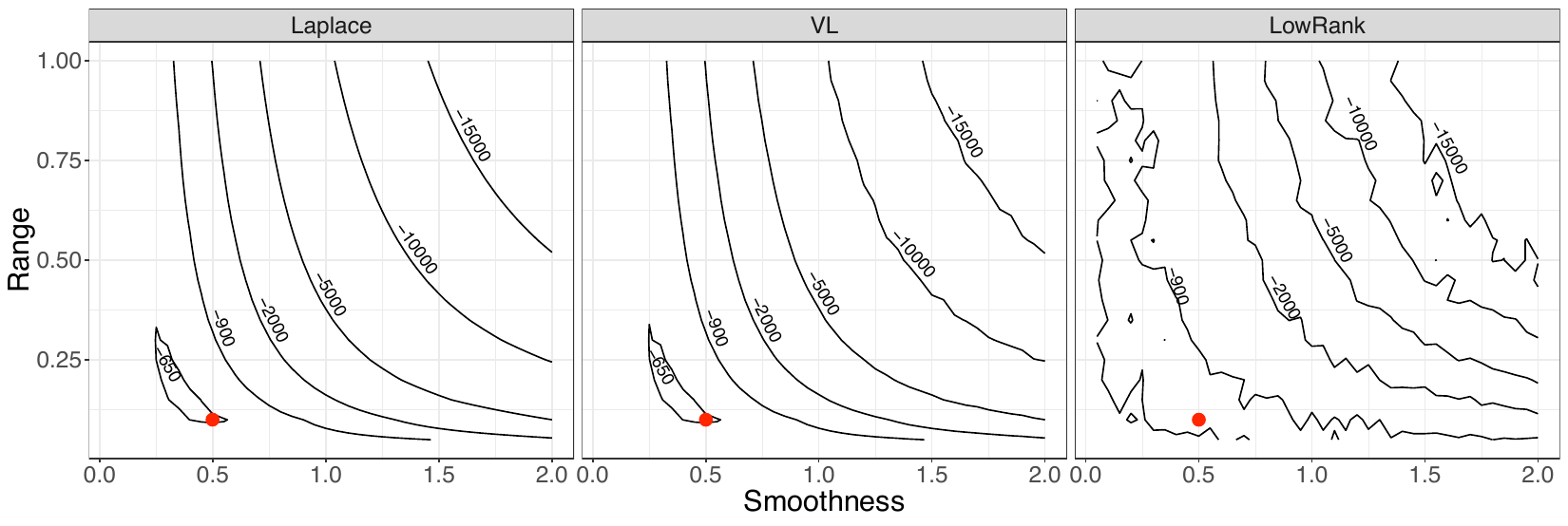}
\caption{For a single realization of Poisson data, contour lines of the integrated likelihood at $(-650, -900, -2000, -5000, -10000, -15000)$}
	\end{subfigure}%

\vspace{3mm}

	\begin{subfigure}{1\textwidth}
	\centering
	\includegraphics[width =.95\linewidth]{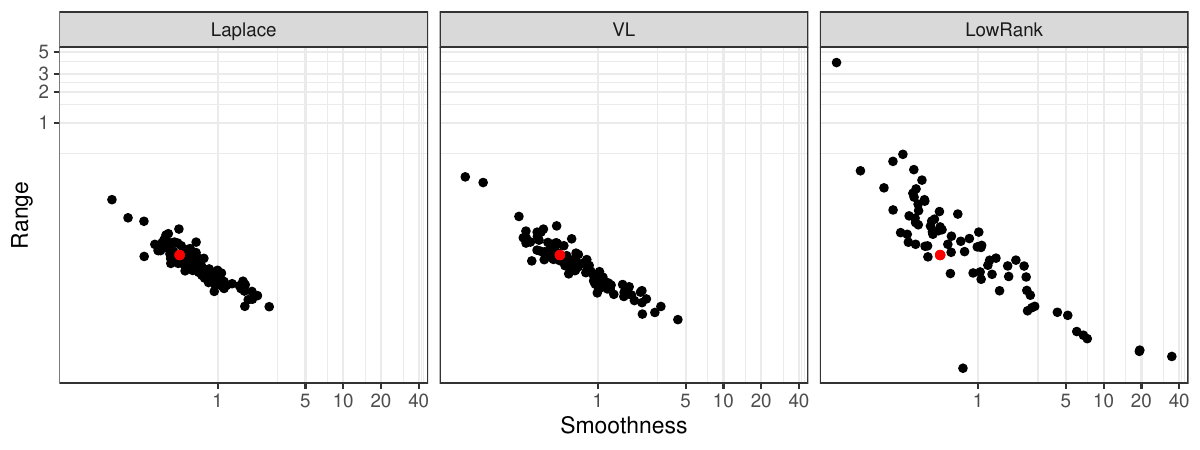}
	\caption{Parameter estimates obtained by optimizing the integrated likelihood for 100 sample realizations; also see Table \ref{tab:param_est_range}}
	\end{subfigure}
 \caption{For Poisson data at $n=625$ locations in the unit square, comparison of different approximations to the integrated likelihood, using conditioning sets of size $m=20$ for VL and LowRank. Red dots show the true parameter values.}
 \label{fig:paramest2d}
\end{figure}

First, we simulated a single realization of the spatial data. Holding the variance fixed at the true value of one, we sequentially evaluated the integrated likelihood on a grid of values for the range and smoothness parameters, using the Laplace approximation in \eqref{eq:laplacelikelihood}, and the VL-RF approximation with $m=20$ in \eqref{eq:vllikelihood}. The exact integrated likelihood is intractable.
As shown in Figure \ref{fig:paramest2d}, the integrated likelihoods as approximated by Laplace and by VL were almost identical, while the LowRank approximation was quite poor. These likelihood approximations are equivalent to approximations to the posterior distribution $\dens(\bftheta|\bz)$ assuming flat priors for $\bftheta$. This indicates that Bayesian inference for GGPs can be carried out quickly and accurately using the VL approximation.

We then simulated 100 different realizations of the spatial Poisson data and examined the parameter estimates obtained by maximizing the different approximations to the integrated likelihood. The scatter plot in Figure \ref{fig:paramest2d} shows the parameter estimates, using $m=20$ conditioning points for VL and LowRank. While the estimates using Laplace and VL were similar, LowRank had significant outliers that increased the RMSE of the parameter estimates (see Table \ref{tab:param_est_range}).  The LowRank parameter estimation frequently diverged due to the rough likelihood surface, and for those cases we repeated the optimization with bounds $(0.001, 20)$ for both range and smoothness, but LowRank still failed repeatedly.

\begin{table}
\centering
\begin{tabular}{l | rrr| r S[table-format=3.2] S[table-format=3.2]|} 
\multirow{2}{*}{} & \multicolumn{3}{c|}{ Range } & \multicolumn{3}{c|}{  Smoothness}  			\\
  &		$m=10$ &  $m=20$ & $m=40$ &		$m=10$ &  $m$\text{=20} & $m$\text{=40}\\
\hline
LowRank  	& 0.107     &  0.407    & 0.098     &   9.17  	&  8.40 	&   12.60\\
VL  	& 0.293     &  0.040 	& 0.023     &  1.01 	&  0.78     &  	0.47 \\ 
\hline
Laplace &           &    &  0.023	            &           &      &  0.51 \\ 
\end{tabular}
\caption{For 100 simulated Poisson datasets at $n=625$ locations in the unit square, RMSE for parameter estimates based on different approximations to the integrated likelihood. Both range and smoothness parameters were bounded to the interval $[0.001, 20]$, but LowRank estimation still failed repeatedly.}
\label{tab:param_est_range}
\end{table}

%%%%
\subsection{Interpretation of simulation results}

In our simulations, VL provided similar accuracy as Laplace with a considerably smaller number $m$ of conditioning points compared to LowRank. The time required per iteration for VL approaches is $\order(nm^3)$. At the expense of fully parallel computation, LowRank can be carried out in $\order(nm^2)$ time by computing the decomposition of the covariance of the conditioning set once at the beginning of the procedure. However, as VL with any given $m$, say $m=\widetilde m$, was substantially more accurate than LowRank with $m = \widetilde m^{3/2}$, we conclude that VL is more computationally efficient than LowRank for a given approximation accuracy, except for very smooth posteriors. The improvement in accuracy for VL relative to LowRank became even more pronounced as we increased the sample size under in-fill asymptotics.

%%%%%%%%%%%%%%%%%%%%%%%%%%%%%%%%%%%%%%%%%%%%%%%%%%%%%%%
\section{Application to satellite data \label{sec:application}}

%\subsection{Data description}

We applied our methodology to a large, spatially correlated, non-Gaussian dataset of column water vapor.  These data were collected by NASA's Moderate Resolution Imaging Spectroradiometer (MODIS), which is mounted on the NASA Aqua satellite \citep{Borbas2017}. We considered a Level-2 near-infrared dataset of total precipitable water at a $1354 \times 2030 = 2{,}746{,}820$ grid of 1km pixels. We used up to 500,000 of these data points for our demonstration.  Our dataset was observed between 13:45 and 13:50 on March 28, 2019 over a rectangular region off the coast of west Africa with west, north, east, and south bounding coordinates -42.707, 67.476, 4.443, and 45.126, respectively and was found on the NASA Earthdata website, \url{https://earthdata.nasa.gov}. 
 
Precipitable water amounts are continuous and strictly positive, with values near 0 corresponding to clear skies and larger values implying more water.  Exploratory plots showed a right-skewed density, so we assumed that the data can be modeled using a spatial generalized GP with a Gamma likelihood:
\[
z(\bs_i)|y(\bs_i) \stackrel{ind.}{\sim} \mathcal{G}(a, ae^{-y(\bs_i)}), \qquad y(\cdot) \sim \mathcal{N}(\mu, K),
\]
where $E(z(\bs)|y(\bs))=\exp(y(\bs))$, $\mu(\bs) = \beta_1 + \beta_2\, \text{lat}(\bs)$ is a linear trend consisting of an intercept and a latitudinal gradient, and $K$ is an isotropic Mat\'ern covariance function with variance $\sigma^2$, smoothness $\nu$, and range parameter $\rho$. 
We estimated the parameter values $\beta_1 = -1.515$, $\beta_2 = 0.000766$, $a = 0.89$, $\sigma^2 = .25$, $\rho = 31$km, and $\nu = 3$ as described in Section \ref{supp:MODIS}. 

%\subsection{Posterior estimation and Predictive accuracy}

We again compared our VL approach to a LowRank method. We randomly sampled $n=250{,}000$ observations $\bz$ of the full dataset as training data, and $250{,}000$ of the remaining observations as test data $\bz^\star$ at locations $\locs^\star$. For VL, we set $m=20$ following our recommendations in Section \ref{sec:2dsim} and further justified in Section \ref{supp:MODIS}. For LowRank, we used $m=89 \approx (20)^{3/2}$ for a computationally fair comparison. On an Intel Xeon E5-2690 CPU with 64GB RAM, Algorithm \ref{alg:vl} for VL required 10 iterations with a total run time of about 18 minutes (1.8 minutes per iteration). Taking advantage of an implementation that achieves the $\order(nm^2)$ scaling, each iteration for LowRank required 1.3 minutes on average across 6 iterations. Note that, based on our numerical experiments, we estimate that Laplace without further approximation would take months of computing time, while HMC-based approaches would take years to achieve the same accuracy as VL.

\begin{figure}
\centering
	\begin{subfigure}{1\textwidth}
	\centering
	\includegraphics[width =.95\linewidth]{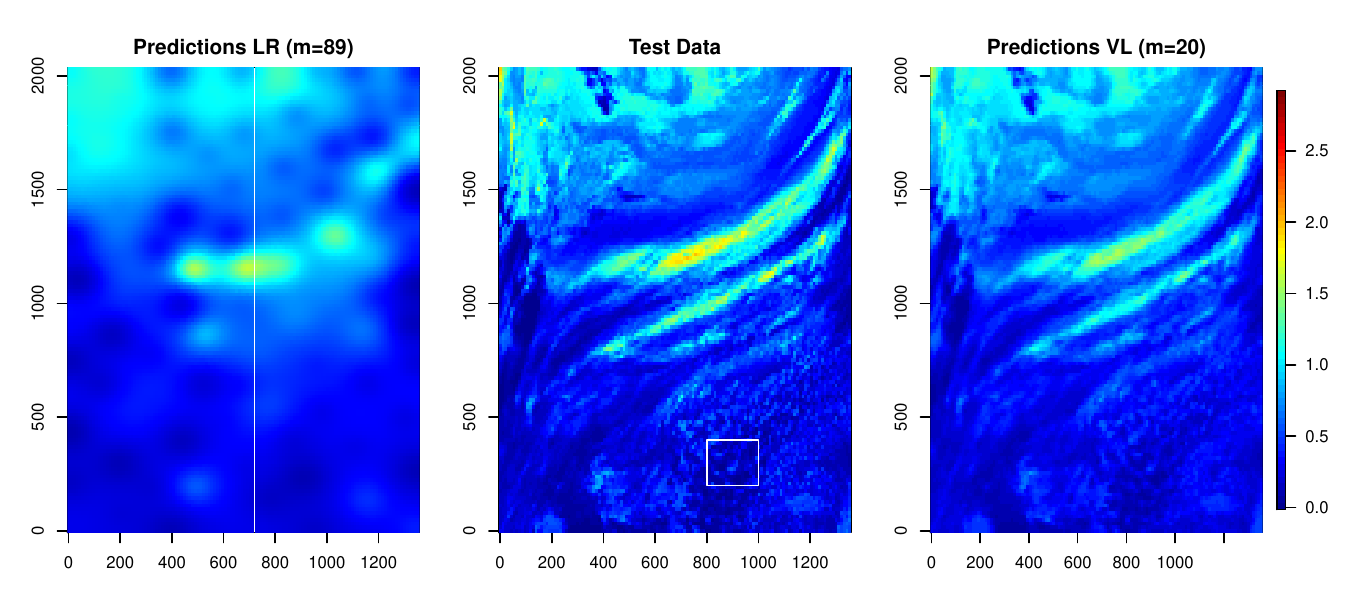}
	\caption{Entire spatial domain}
	\label{fig:map_all}
	\end{subfigure}%

\vspace{3mm}

	\begin{subfigure}{1\textwidth}
	\centering
	\includegraphics[width =.93\linewidth]{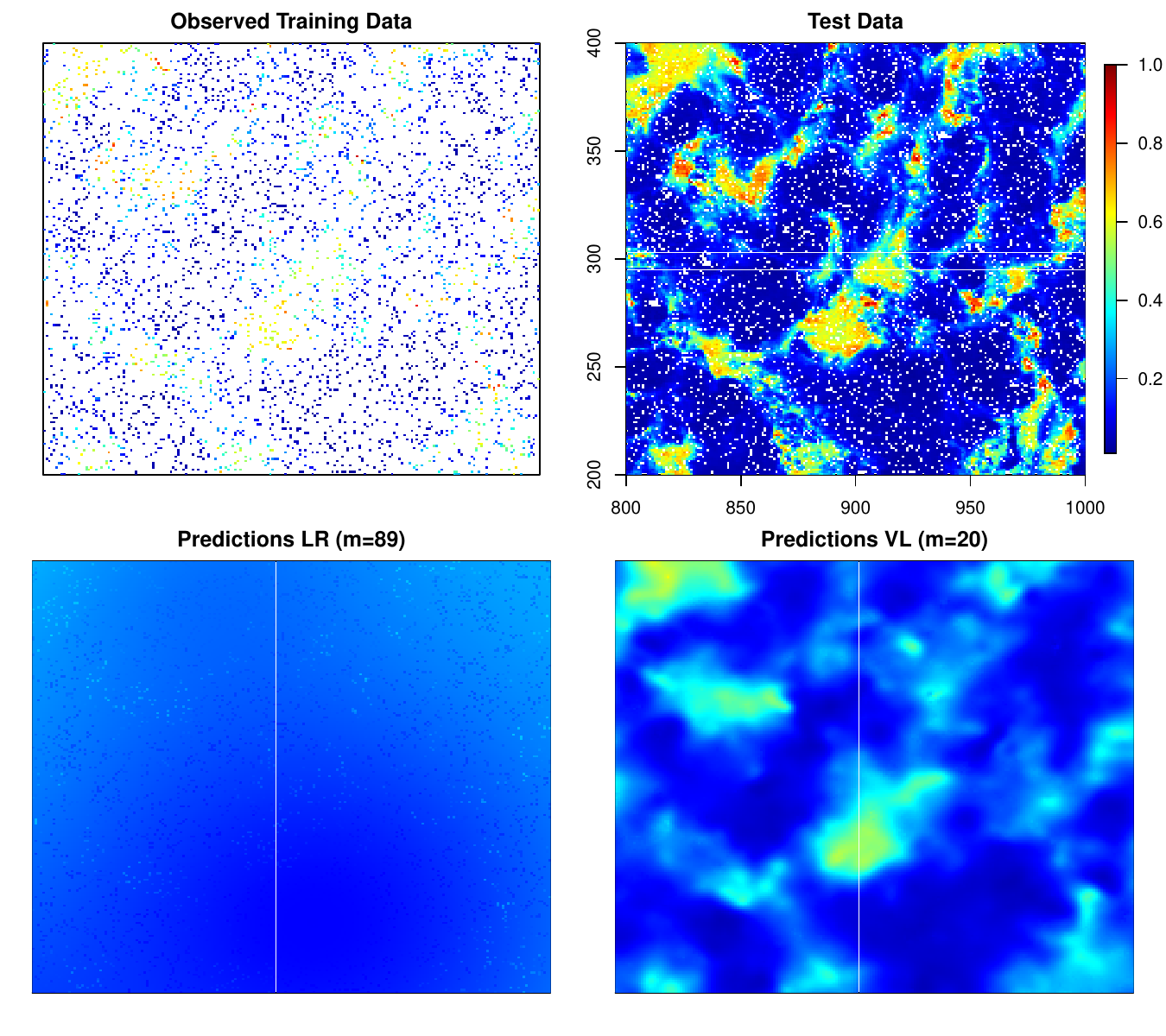}
	\caption{Zooming into the white square shown in Panel (a)}
	\label{fig:map_zoom}
	\end{subfigure}
 \caption{Prediction maps for MODIS data using VL and LowRank (LR)}
 \label{fig:MODIS}
\end{figure}

Figure \ref{fig:map_all} shows prediction maps of the posterior mean $E(\bz^\star|\bz) = E(\exp(\by^\star)|\bz)$ with $i$th entry $\exp(E(y_i^\star|\bz)+var(y_i^\star|\bz)/2)$. Clearly, much of the fine-scale structure was lost when using LowRank. To further illustrate this issue, we made predictions on a $200\times 200$ grid over a small subregion. As shown in Figure \ref{fig:map_zoom}, the LowRank predictions were virtually useless at this scale, while VL was able to recover much of the important spatial structure from the noisy and incomplete training data.

Table \ref{tab:MODIS_scores} quantifies the improvement in predictions using VL over LowRank. We computed the MSE based on the posterior mean $E(\bz^\star|\bz)$. To compare the accuracy of the uncertainty quantification, we also computed the continuous ranked probability score \citep[CRPS; e.g.,][]{gneiting2014probabilistic}, which encourages well calibrated and sharp predictive distributions. Table \ref{tab:MODIS_scores} shows that VL strongly outperformed LowRank for comparable computational complexity.

\begin{table}
\centering
 \begin{tabular}{l | r | r} 
Method & MSE & CRPS \\ 
 \hline
 VL & 0.0149 &  0.144  \\ 
 LowRank  & 0.0528 & 0.170  \\ 
 \hline
 Ratio & 3.54$\times$ &  1.18$\times$\\
\end{tabular}
\caption{For the MODIS data, comparison of prediction scores (lower is better) between VL and LowRank}
\label{tab:MODIS_scores}
\end{table}

%%%%%%%%%%%%%%%%%%%%%%%%%%%%%%%%%%%%%%%%%%%%%%%%%%%%%%%
\section{Conclusions and future work}

In this work, we presented a novel combination of techniques that allow for efficient analysis of large, spatially correlated, non-Gaussian datasets or point patterns. The key idea is to apply a Vecchia approximation to the Gaussian (and hence tractable) joint distribution of GP realizations and pseudo-data at each iteration of a Newton-Raphson algorithm, leading to a Gaussian Laplace approximation. This allows us to carry out inference for non-Gaussian data by iteratively applying existing Vecchia approximations for Gaussian pseudo-data, which are updated at each iteration. Our Vecchia-Laplace (VL) techniques guarantee linear complexity in the data size while capturing spatial dependence at all scales. Compared to alternative methods such as low-rank approximations or sampling-based approaches, our VL approximations can achieve higher accuracy at a fraction of the computation time.

Vecchia approximations require specification of an ordering of the model variables and of a conditioning set for each variable, and these two issues also play a critical role in the performance of our VL approaches. Through simulation studies, we showed that, in one-dimensional space, interweaving the GP realizations and the pseudo-data \citep{Katzfuss2017a} can provide results that are virtually indistinguishable from Laplace, even for very small conditioning sets. For two-dimensional space, we recommend the response-first Vecchia approximation \citep{Katzfuss2018}. Due to the computational efficiency of our approach, it is also possible to use a VL approximation of the integrated likelihood for parameter inference, for which we recommend the interweaved ordering in any dimension.

The methods and algorithms proposed here are implemented in the \texttt{R} package \texttt{GPvecchia} \citep{GPvecchia}. The default settings of the package functions reflect the recommendations in the previous paragraph. The tuning parameter $m$, which controls a trade-off between accuracy and computation cost, can be set by the user. In practice, a useful strategy is to start with a relatively small value of $m$ and gradually increase it until the inference converges or the computational resources are exhausted.

Our methods and code are applicable in more than two dimensions, but a thorough investigation of their properties in this context will be carried out in future work. For example, \citet{Katzfuss2020} show that Vecchia-based approximations with appropriate extensions can be highly accurate for computer-model emulation in up to ten dimensions; a combination with our VL methods could allow emulation of non-Gaussian computer-model output.
Other potential future work includes extending the Laplace approximation in our methods to an integrated nested Laplace approximation (INLA) that improves the accuracy of the marginal posteriors of the latent variables \citep[][Sect.~3.2]{Rue2009}; the use of conjugate-gradient \citep{zhang2019practical} or incomplete-Cholesky \citep{Schafer2020} methods that allow the computation of the latent posterior mean in linear time even for completely latent Vecchia approximations; or extensions to spatio-temporal filtering using Vecchia approximations based on domain partitioning \citep{Jurek2018,Jurek2020}.

%%%%%%%%%%%%%%%%%%%%%%%%%%%%%%%%%%%%%%%%%%%%%%%%%%%%%%%
\footnotesize
\appendix
\section*{Acknowledgments}

Katzfuss' research was partially supported by National Science Foundation (NSF) grant DMS--1521676 and NSF CAREER grant DMS--1654083. The authors would like to thank Joe Guinness, Jianhua Huang, Leonid Koralov, Boris Vainberg, and several anonymous reviewers for helpful comments and suggestions. Wenlong Gong, Joe Guinness, Marcin Jurek, and Jingjie Zhang contributed to the R package \texttt{GPvecchia}, and Florian Sch\"{a}fer provided C code for the exact maxmin ordering.

%%%%%%%%%%%%%%%%%%%%%%%%%%%%%%%%%%%%%%%%%%%%%%%%%%%%%%%

\section{Newton-Raphson update using pseudo-data \label{app:nr}}

The desired Newton-Raphson update has the form
\begin{equation}
\label{eq:nrupdate}
\textstyle \bh(\by) = \by - \big(\frac{\partial^2}{\partial \by\by'} \log p(\by|\bz)\big)^{-1}\big(\frac{\partial}{\partial \by} \log p(\by|\bz)\big).
\end{equation}
As shown in Section \ref{sec:laplace}, we have $\frac{\partial}{\partial \by} \log p(\by|\bz) = \bK^{-1}(\bfmu-\by) + \bu_\by$ and $-\frac{\partial^2}{\partial \by\by'} \log p(\by|\bz) = \bK^{-1} + \bD_\by^{-1} = \bW_\by$.  Using an idea similar to iterative weighted least squares \citep[Section 2.5,][]{mccullagh1989generalized}, we can premultiply the variable $\by$ by the Hessian to combine terms, and then rearrange and pull out the prior mean.  Dropping the iteration subscript of $\by$ for ease of notation, we can write \eqref{eq:nrupdate} as
\begin{align*}
\bh(\by) & = \by + \bW^{-1}(\bK^{-1}(\bfmu - \by) + \bu) \\
 & = \bW^{-1}\big((\bK^{-1} + \bD^{-1})\by - \bK^{-1}\by + (\bK^{-1}\bfmu + \bD^{-1}\bfmu) - \bD^{-1}\bfmu + \bD^{-1}\bD\bu \big)\\
 & = \bfmu + \bW^{-1}\big(\bD^{-1}(\by + \bD\bu - \bfmu)\big)\\
 & = \bfmu + \bW^{-1}\bD^{-1}(\bt - \bfmu), 
\end{align*}
where $\bt = \by + \bD \bu$.

Now consider the posterior mean in the case of a Gaussian likelihood $\bt|\by \sim \normal_n(\by,\bD)$ with a conjugate Gaussian prior, $\by \sim \normal_n(\bfmu,\bK)$. Employing a well-known formula, we have
\[
\E(\by|\bt) = (\bK^{-1} + \bD^{-1})^{-1}(\bK^{-1}\bfmu + \bD^{-1}\bt) = \bfmu + \bW^{-1}\bD^{-1}(\bt - \bfmu).
\]
Thus, we have $\bh(\by) = \E(\by|\bt)$, the posterior mean under the assumption of Gaussian pseudo-data $\bt$.

%%%%
\section{Computing $\bU$ \label{app:computeU}}

Consider a general Vecchia approximation of the form \eqref{eq:vecchia}. To obtain $\bU$, define $C(x_i,x_j)$ as the covariance between $x_i$ and $x_j$ implied by the true model; that is, $C(y_i,y_j) = C(t_i,y_j) = K(\bs_i,\bs_j)$ and $C(t_i,t_j) = K(\bs_i,\bs_j) + \indicat_{i=j} d_i$. Then, the $(j,i)$th element of $\bU$ can be calculated as
\begin{equation}
\label{eq:U}
\bU_{ji} = \begin{cases} r_i^{-1/2}, & i=j,\\ -b_{i}^{(j)} r_i^{-1/2}, & j \in c(i), \\ 0, &\textnormal{otherwise}, \end{cases}
\end{equation}
where $\bb_i'= C(x_i,\bx_{c(i)}) C(\bx_{c(i)},\bx_{c(i)})^{-1}$, $r_i = C(x_i,x_i) - \bb_i' C(\bx_{c(i)},x_i)$, and $b_i^{(j)}$ denotes the $k$th element of $\bb_i$ if $j$ is the $k$th element in $c(i)$ (i.e., $b_i^{(j)}$ is the element of $\bb_i$ corresponding to $x_j$).

%%%%%%%%%%%
\footnotesize
\bibliographystyle{apalike}
\bibliography{library}

\end{document}